# Practical Adoption of Cloud Computing in Power Systems – Drivers, Challenges, Guidance, and Real-world Use Cases

IEEE Task Force on Cloud Computing for Power Grid

Song Zhang (TF Chair), *Senior Member, IEEE*, Amritanshu Pandey (TF Vice-Chair), *Member, IEEE*, Xiaochuan Luo (TF Vice-Chair), *Senior Member, IEEE,* Maggy Powell, *Member, IEEE,* Ranjan Banerji, *Member, IEEE,* Lei Fan*, Senior Member, IEEE,* Abhineet Parchure, *Member, IEEE*, Edgardo Luzcando, *Member, IEEE*

*Abstract*—Motivated by the Federal Energy Regulatory Commission's (FERC) recent direction and ever-growing interest in cloud adoption by power utilities, a Task Force was established to assist power system practitioners with secure, reliable and cost-effective adoption of cloud technology to meet various business needs. This paper summarizes the business drivers, challenges, guidance, and best practices for cloud adoption in power systems from the Task Force's perspective, after extensive review and deliberation by its members, including grid operators, utility companies, software vendors, and cloud providers. The paper begins by enumerating various business drivers for cloud adoption in the power industry. It follows with the discussion of the challenges and risks of migrating power grid utility workloads to the cloud. Next, for each corresponding challenge or risk, the paper provides appropriate guidance. Notably, the guidance is directed toward power industry professionals who are considering cloud solutions and are yet hesitant about the practical execution. Finally, to tie all the sections together, the paper documents various real-world use cases of cloud technology in the power system domain, which both the power industry practitioners and software vendors can look toward to design and select their own future cloud solutions. We hope that the information in this paper will serve as helpful guidance for the development of NERC guidelines and standards relevant to cloud adoption in the industry.

## Key Terminologies

*IT/OT* – Information Technology and Operational Technology

*Infrastructure-as-a-Service* – a category of cloud computing services that provide essential virtualized resources such as computing, storage and networking over the Internet

*Platform-as-a-Service* – a category of cloud computing services that allows customers to provision, instantiate, run, and use a modular bundle comprising of a computing platform and one or more applications without the need for users to perform essential management (e.g., patching)

*Software-as-a-Service* – a category of cloud computing services where a vendor hosts applications and deliver them to end-users over the Internet on a subscription basis

*Function-as-a-Service* – a category of cloud computing services that allow users to develop, deploy and run single-purpose applications by modules without having to manage servers

*Container-as-a-Service* – a category of cloud computing services that provides software developers and IT departments to upload, organize, run, scale, and manage containers by using container-based virtualization

*Cloud Service Provider* – a company that offers components of cloud computing such as the services mentioned above (e.g., container service)

*Elastic Computing* – a cloud computing characteristic that denotes secure and resizable compute capacity to meet changing demands without fixed preplanning for capacity and engineering for peak usage

*Serverless Computing* – a cloud computing execution model in which the cloud provider provides an execution environment that does not require users to manage servers

*Virtual Machine* – virtualization of a computer system.

S. Zhang and X. Luo are with ISO New England, Holyoke, MA 01040 USA (e-mail: sozhang@iso-ne.com, xluo@iso-ne.com).

A. Pandey was with Pearl Street Technologies, Pittsburgh, PA, USA. He is now with the Department of ECE, Carnegie Mellon University, Pittsburgh, PA 80523 USA (e-mail: amritanp@andrew.cmu.edu).

M. Powell, R. Banerji, and A. Parchure are with Amazon Web Services, Herndon, VA 20170, USA (e-mail: maggyp@amazon.com, rbanerji@amazon.com, parchua@amazon.com).

L. Fan is with University of Houston, Houston, TX 77204, USA (e-mail: lfan8@Central.UH.EDU).

E. Luzcando was with Midcontinent ISO from 2003 to 2012 and New York ISO from 2012 to 2018. He is now with Performance Improvement Partners, Shelton, CT 06484, USA (e-mail: eluzcando@pipartners.com).

Virtual machines use the hypervisor to share and manage hardware, allowing for multiple environments which are isolated from one another on the same physical machine

*Vertical Scaling (Scaling Up)* – scaling by adding more resources to a single node such as additional CPU, RAM, and DISK to cope with an increasing workload

*Horizontal Scaling (Scaling Out)* – scaling and increasing resources by adding more nodes (e.g., virtual machines) to an existing pool of nodes rather than simply adding resources to a single node

*Regions and Availability Zones* – Regions are geographic locations in which cloud service providers data centers are geographically located. Availability Zones are isolated locations within data center Regions from which cloud services originate and operate;

*Shared Responsibility Model* – a cloud security framework that dictates the security obligations of a cloud computing provider and its users to ensure accountability

*Spot instance* – a node (e.g., virtual machine) that uses a spot market pricing model for a CSP's unused capacity. Spot instances come at a steep discount but can be shut down when the CSP no longer has unused capacity.

*On-Demand Instance*– a node (e.g., virtual machine) that you pay for by the hour or second with no long-term commitments.

*Reserved Instance*– a node (e.g., virtual machine) that you pay a discounted rate for by reserving compute capacity and committing to long-term usage (for example, 1 to 3 years).

I. INTRODUCTION

Cloud computing is a mature technology, which has modernized many business enterprises. It is globally seen as a critical infrastructure [1] like other vital resources such as power, gas, and freshwater supply [2]. A recent report [3] found that 99% of enterprises gained significant technical benefits from adopting cloud technology and 77% of total enterprises used public cloud in some capacity. The same report found that 81% of the enterprises believed that they could innovate more quickly while working on the cloud than on-premise infrastructure. Cloud technology has significant benefits over other traditional forms of computing and data storage and analysis.

Given the significant benefits of cloud technology, the power industry is likely to adopt cloud as well. However, the adoption of cloud technology in the power industry faces resistance on several fronts, e.g., cyber security, compliance, cost, consistency, latency, software pricing and licensing. Nonetheless, there is a growing interest in cloud adoption amongst many grid entities driven by their business needs. This is partly driven by rapid grid modernization and decarbonization, which requires ever-growing demand for data analytics and resources such as computing, network, and storage. Traditional on-premises facilities face constraints, and the most affected utilities are eagerly searching for scalable and cost-effective solutions to meet their fast-rising needs. Cloud technology is an obvious choice. Power system practitioners are continuously seeking advanced algorithms and new solution frameworks that can benefit from elastic compute resources inherent in cloud computing. This, along with cost-effective storage options, makes cloud technology an ideal option for power system practitioners.

Cloud technology has other benefits as well. The modern power grid is a cyber-physical system that integrates the physical grid infrastructure with the Information Technology/Operational Technology (IT/OT) infrastructure for reliable and resilient grid design and operation. Such safety-critical systems should have a fault-tolerant architecture to ensure operational continuity in IT/OT systems disruption [4] [5] [6] [7]. Cloud computing, a proven technology in several other industries such as finance, e-commerce, insurance, and healthcare [8], is a fitting option for these grid applications because it offers fault-tolerant system design capabilities and benefits without an equal (linear) increase in cost. Learning from numerous successful applications in the industries mentioned above that impose no less stringent requirements on cyber security and compliance, the Task Force concludes that cloud technology can also benefit the entire power industry if it is adopted securely and reliably.

Despite the power industry's firm resistance to cloud adoption often due to misunderstood fundamentals, some innovative organizations have begun to use the cloud for non-critical, low-impact workloads such as planning studies and load forecasting. Although there has been some previously published work, such as [9] [10] [11], none have surveyed the application of cloud technology in the power industry comprehensively and practically. [9] discusses cloud only from the security perspective; [10] enumerates a few application scenarios but provides only one private cloud-based use case and does not include discussion of the challenges and solutions for cloud adoption. [11] summarizes the technical issues and possible solutions for the cloud, but as a general literature review paper rather than for the power industry. Given this background, the Task Force initiated this work to systematically organize the most recent cloud applications in the electric energy sector and provide expert guidance for challenges in cloud adoption by power system users and software vendors. To provide the guidance, we combine i) best practices endorsed by the Cloud Service Providers (CSP) and ii) experiences learned from the pioneering use cases of cloud technology in the power industry. This work aims to address the common concerns over cloud adoption, provide guidance for reliable and secure use of cloud resources for power entities, and elaborate on how cloud technology can help in various power system businesses. Additionally, this document also aims to untangle common misconceptions related to cloud technology in the power industry and help software vendors design products that are better suited to the cloud.

*The authors of the paper recommend that the readers consider the following before reading the document. The paper is divided into four independent broad sections. Section II first gives a brief introduction to cloud computing basics and then describes the key business drivers for cloud adoption in the field of power systems. If the reader's interest lies solely on drivers*

*for cloud adoption, they should read Section II. Section II summarizes a few typical real-world examples of cloud adoption in the power industry that correspond to many business drivers. Section III of the paper documents the known challenges from the power grid utilities and software vendors' perspective when adopting or developing cloud technology. Section IV is closely tied to Section III and provides guidance to the reader for every challenge documented in Section III. Suppose the reader is interested in reviewing a specific challenge. In that case, they can jump directly to the relevant sub-section in Section III and follow that with corresponding guidance for that challenge in Section IV. Finally, Section V brings together real-world use cases of cloud technology in power grids, with details provided in a preprint of this work.*

## II. Cloud Fundamentals and Business Drivers For Cloud Adoption

### A. Cloud Computing Basics

Cloud computing can be categorized into public, private, and hybrid clouds based on their ownership.

A private cloud consists of cloud computing resources that are used solely by one enterprise or organization. The private cloud can be physically located at an organization's on-campus data center or hosted by a third-party service provider. In a private cloud, the services are always maintained on a private network, and the infrastructure, hardware and software belong exclusively to the organization. A private cloud makes it easier for an organization to customize its resources to meet specific IT requirements compared to other cloud types. Historically, private clouds were often used by organizations with business-critical operations seeking enhanced control over their environment, such as government agencies, financial institutions and healthcare companies. But in recent years, the creation of cloud-friendly compliance requirements such as FedRAMP for the federal government, HIPAA and HITRUST CSF for the healthcare industry, PCI and SEC Rule 17a-4(f) for financial services [12] [13] have opened the door to the adoption of public cloud by critical sectors of the economy.

Public clouds are the most common type of cloud computing deployment. The resources leased by users, including all hardware, software, and other supporting infrastructure for computing, storage, networking, etc., are owned and operated by CSPs and delivered over the Internet. The primary public cloud providers are Amazon Web Services, Microsoft Azure, Oracle Cloud, Google Cloud, Salesforce, IBM, RedHat, Alibaba, and Tencent.

A hybrid cloud combines on-premises infrastructure—or a private cloud—with one or more public cloud services. Hybrid clouds allow data and applications to move between the two environments. Many organizations choose a hybrid cloud approach due to many business imperatives. These include meeting regulatory and data sovereignty requirements, reducing network latencies, taking advantage of on-premises technology investment while simultaneously maintaining the ability to scale to the public cloud and paying for extra computing power only when needed.

From the service model perspective, cloud computing is typically delivered to customers in terms of Infrastructure-as-a-Service (IaaS), Platform-as-a-Service (PaaS), and Software-as-a-Service (SaaS). In recent years, other service models such as Container-as-a-Service (CaaS) and Function-as-a-Service (FaaS) are also emerging as the new attractive cloud offerings. Due to the page limit, we only introduce the three most common service models: IaaS, PaaS and SaaS.

*SaaS* - While the SaaS model gives the least control over the software and the underlying services, it also provides significant benefits on other fronts. SaaS is affordable as it eliminates the costs involved with purchasing, installing, maintaining, and upgrading computing hardware. With SaaS, the services can be accessed from any device such as company laptops, smartphones, and tablets, eliminating the constraints set by on-premises software. SaaS has been incorporated into the business strategy of nearly all enterprise software companies. For example, Siemens PTI and Energy Exemplar have made the cloud version of their software to attract customers.

*IaaS* - IaaS is the most flexible service model that gives the best control over the hardware infrastructure, such as managing identity and access, customizing guest operating systems and upper-level applications according to the users' requirements. Deploying an IaaS cloud model eliminates the need to deploy on-premises hardware, which helps to reduce the total costs of ownership.

*PaaS* - Compared to IaaS and SaaS, PaaS is the middle layer where you can offload most of the work to the provider and fill in the gaps as needed. PaaS reduces the development time since the vendor provides all the essential computing and networking resources, simplifying the process and improving the development team's focus.

For more details about various cloud service models, please refer to Chapter 1 of [14].

### B. Advantages of Cloud Technology over On-premises Technologies

Cloud techonolgy has several advantages over other on-premises technologies as listed in Table I. These advantages apply to many sub-technologies, which are available under both cloud and on-premises computing such as high-performance computing, distributed storage, distributed computing, etc. The attributes of cloud technology that will benefit the power industry include but are not limited to always-on availability, elastic capacity, massive and scalable data storage, improved collaboration, excellent accessibility, and low maintenance cost [15].

TABLE I
CLOUD ADVANTAGES OVER ON-PREMISES TECHNOLOGIES

| Advantage | What does it mean? |
|---|---|
| Agility | Quickly create/edit infrastructure thus enabling frequent experimentation and innovation |
| Cost Savings | Only pay for what you use, lower upfront expenses |
| Resilience | Highly available across multiple regions and automation capabilities to build and recover from failure/disasters |
| Elasticity | Easily Scale up or down with the needs of the business |
| Innovate Faster | Ability to focus on business differentiators, not infrastructure |
| Go Global in Minutes | Use CSP provided tools for agility and their global presence to provide services around the world |

*C. Business Drivers for Cloud Adoption in the Power Industry*

The advent of cloud computing has brought unprecedented benefits to organizations in many business sectors, and the power industry is no exception. Digital transformation of the electric grid is an essential catalyst for cloud adoption. Broadly the business drivers for cloud technology in the power industry can be viewed from three different perspectives – *resources*, *solutions*, and *infrastructure*.

*A resources viewpoint:* The modern power grid requires flexible access to IT resources that are scalable and cost-effective. These resources include but are not limited to storage, computing, and networking. Due to large-scale deployments of new OT systems such as Phasor Measurement Unit (PMU), Advanced Metering Infrastructure (AMI), smart meters, and industrial Internet of Things (IoT) devices, the data generated by the daily power system activities has exploded in the past decade. As with other industries, this data is a "gold mine" for improved analysis, operation, predictive maintenance, planning, monitoring and control; the need for which has become more imminent due to uncertainty and variability in operating patterns of the grid. These are primarily driven by the proliferation of renewable energy and distributed resources, risks associated with extreme weather events, ambitious goals to shift from fossil fuels to low-carbon technologies, and rapid electrification. As one would expect, the efficacy of these improved analyses depends on the storage, management and processing of large volumes of data collected in the grid OT environment. How to efficiently manage and process such data and uncover the value behind them is a question many grid utilities face today. Cloud computing is a practical and economical option for the power industry to acquire massively scalable and elastic resources for data transmission, storage, processing and visualization. In only a few years, CSPs offerings have rapidly evolved, from compute technologies that started with virtual machines to serverless computing [16] containers, and enhanced orchestration (Kubernetes). All the offerings enable a flexible way to acquire resources without a complex and time-consuming procurement process (i.e., "pay-as-you-go" pricing strategy). Utility companies only need to pay for the time period when they utilize the resources *leased* from the CSPs.

*A solutions viewpoint:* Cloud computing unlocks numerous new solution frameworks, advanced algorithms, and tools such as data-driven approaches through data mining, Machine Learning (ML) and Artificial Intelligence (AI), in a highly accessible manner. These data-driven methods can be used to develop online algorithms that had been traditionally difficult for model-based methods due to the lack of suitable model parameters. These algorithms have applications ranging from control to cyber-security (anomaly detection) to weather forecasting. CSPs constantly update and add new technologies (e.g., ML/AI models) that power utilities can immediately test and use to enhance their data analysis capabilities and improve business outcomes. These services based on a "pay-as-you-go" billing approach considerably lower power system users' barrier to utilizing the ML methods and frameworks, which are typically highly demanding for hardware and software. With cloud technology, power utilities can now take advantage of these advanced algorithms and focus on their primary business needs, letting the CSPs do the heavy lifting for them.

*An infrastructure viewpoint:* The IT/OT cyber infrastructure of power system entities, whether on the supply-side or demand side, needs a revamp to better adapt to the challenges brought by grid modernization. For instance, the transmission operators and distribution operators need a resilient IT infrastructure that offers excellent local and geographical redundancy to support operational continuity after blackouts and other extreme weather events like hurricanes, wildfires, earthquakes and tsunamis. CSPs provide such resilience and redundancy by offering services from more than one geographic region, automation for monitoring, rapid reactions to such devastating events, and elasticity to scale as needed. With cloud technology, power utilities can quickly implement a fault-tolerant architecture that can support their business continuity if their infrastructure experiences any disturbances. In addition, the 5G and IoT revolution have a profound impact on many industries, including power systems. To unleash the power of these cutting-edge technologies, power utilities need to respond to the business needs of OT systems such as SCADA and EMS while ensuring security, integration, visibility, control and compatibility. Thus, power utilities need to carefully consider the right approach to bring 5G and IoT devices to the enterprise workload. Cloud technology can simplify the integration of these technologies and unify IT and OT systems to construct a converged system architecture.

Fig. 1 shows a non-exhaustive list of power system business needs that cloud services can support. It covers a wide range of power system entities, including grid operators, distribution utilities and market participants. The following paragraphs show how these application instances (in Fig. 1) can benefit from cloud adoption. Furthermore, in Section V, we also summarize some of the real-world use-cases for some of the application instances outlined within Fig. 1.

*1) Planning studies*

As the power grid continues its modernization journey, its network size and complexity is constantly increasing. More and more sophisticated interconnections include new devices like High Voltage Direct Current (HVDC) systems and Flexible Alternating Current Transmission System (FACTS) and significant penetration of renewables and electrified resources. These add many nonlinear, non-convex and ill-behaved characteristics to the grid analysis. A large set of discrete variables and differential and algebraic equations must be added to the grid analytical model to study such complex features. Adding further to the complexity is the uncertainty and variability brought about by utility-scale interconnected renewables, Distributed Energy Resources (DERs) and microgrids that require modeling many grid parameters as random variables. Together these features have led to engineers having to analyze a far greater range of scenarios to comprehensively evaluate the impact on the system from many different perspectives, including voltage, thermal and stability. This requires an order of magnitude increase in computational resources.

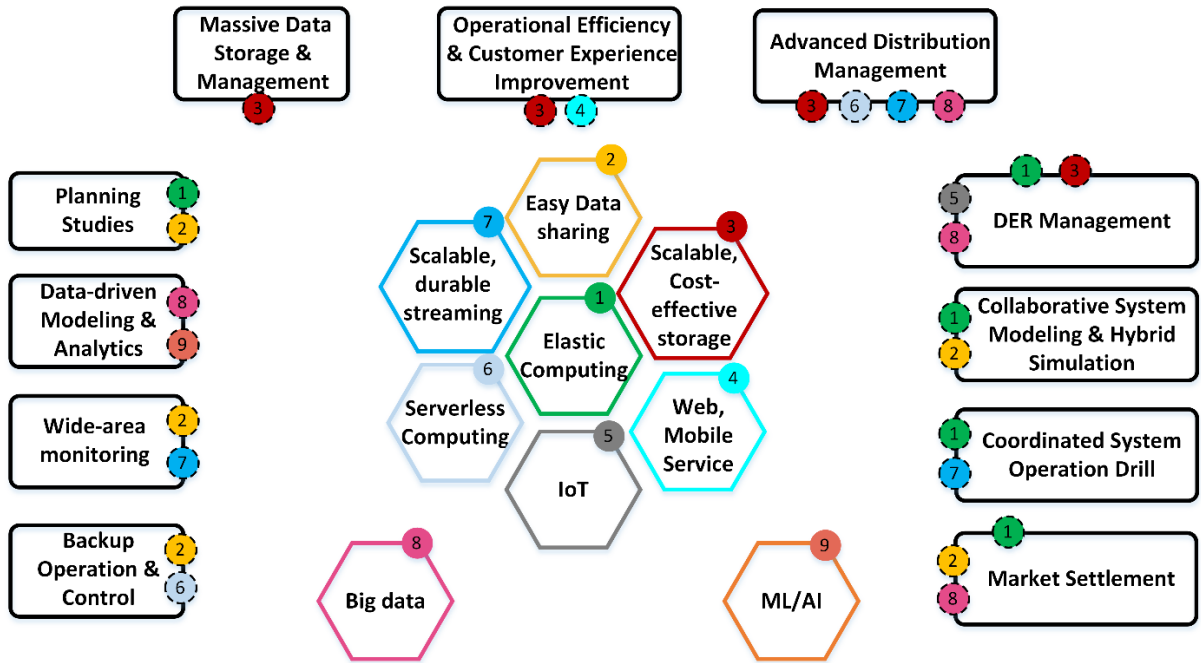

Fig. 1. Mapping the business needs to cloud services. The hexagons in the center of the diagram represent a few categories of cloud services, while the surrounding rectangles represent various business drivers. Numbered and colored circles are attached to distinct rectangles, indicating which business need will primarily benefit from (or rely on) what cloud services/advantages. For example, planning studies mainly take advantage of elastic computing and accessible data sharing features of the cloud.

Despite the need to manage fast-growing computational intensity and complexity, the computing resources provisioned by power system organizations today are not meeting the computational needs of emerging grid analyses, and so-called on-premises "supercomputers" or "computing clusters" are creating computational bottlenecks. These local computing resources are unable to scale and are inelastic. As the demand increases, they constrain the simulation efficiency and underutilize the sunk cost when idle. To meet the peak demand for computation while simultaneously striking a balance between cost and efficiency, elasticity or scalability, is needed imminently by grid operators, transmission owners, and vertically integrated utilities. Elasticity is defined as one of the key feautures of cloud computing by NIST in [17]. Cloud computing embodies HPC and parallel computing and allows the users to rapidly upscale/downscale the resources as needed without worrying about capacity planning. Such a type of cloud service is called elastic computing. The elasticity of cloud resources eliminates the long waiting time from the procurement to the deployment in an on-premises environment. Through the cloud, power system engineers can perform large-scale simulations more cost-effectively by easily scaling out the resources to meet their computing needs. Such supersized studies come in various forms: steady-state analyses, such as transmission needs assessment, installed capacity requirement study and tie benefit study, or dynamic studies such as transient stability assessment and cascading analysis. Moreover, with access to elastic cloud computing, engineers can solve previously unsolvable time-constrained problems, for instance, generating units delist study during forward capacity auctions [18].

It should be noted that elastic computing and on-premises HPC are not mutually exclusive. On the contrary, they complement each other. As shown in Fig. 2, while the cloud resources are the best choice for meeting the variable part of the demand curve considering its rapid elasticity, the on-premises computing resources are a good resource for the fixed portion of demand since the predictable workload can be met by the resources provisioned in-advance. Moreover, the on-premises resources can complement the cloud as a local backup of cloud resources during certain exigent circumstances, e.g., network outages and cloud service outages.

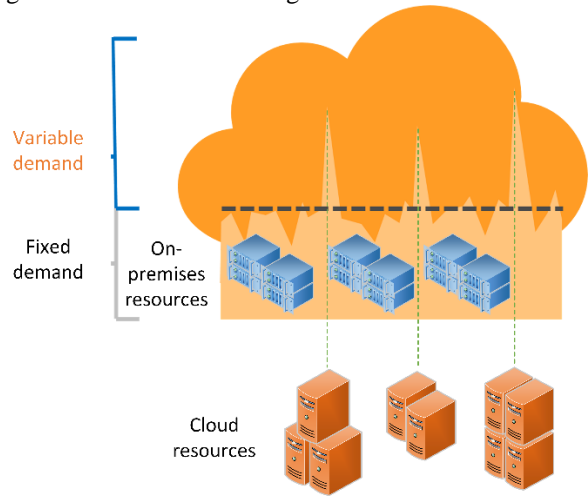

Fig. 2. Using on-premises and cloud computing to meet different portions of demand: the computing demand of a utility can be typically divided into two parts: fixed demand and variable demand. The former indicates the predictable, planned workload, whereas the latter refers to the bursting, unplanned workload.

*2) Storage and Management of Massive System Data*

The demand for massive data storage and management had gradually increased within power system entities, even before the so-called "era of big data" arrived. Generating purchase plans for data storage media has been one of the critical annual tasks undertaken by the IT department in most power system companies since pre-big-data times and more so now. Power utilities are experiencing an exponential increase in yearly intake rates. They are the outposts of most end-user energy data [19]. The growth of this user-generated data is primarily driven by emerging technologies such as AMI smart meters and IoT devices. In the meantime, the deployment of PMU devices across the interconnected bulk power system has also contributed significantly to the tremendous growth in data intake because of the high sampling rate and increasing penetration.

Many industries share these needs, and the power industry is no exception. With the exponentially growing data intake, the power industry seeks a high-speed, low-cost and scalable data storage approach. Cloud technology is an ideal solution for data storage of this scale, where power utilities no longer have to worry about over-provisioning or under-provisioning storage. With this dynamic and almost "unlimited" access to data storage, the cloud offers the underpinning for seamlessly integrated advanced data management. IT administrators and engineers can organize, protect, process, and validate the grid data more conveniently and effectively with these data-related resources and services.

*3) Data-driven Modeling and Analytics*

As the amount of data generation and intake grows, power system companies, like many other industries, are looking to advanced algorithms and approaches to uncover the value behind their data [19]. The data-driven methods, especially ML and AI techniques, offer an excellent approach to achieving these goals.

Although rarely mentioned, efforts to apply traditional AI/ML methods in power systems began a few decades ago. For instance, Artificial Neural Network (ANN) was widely studied in the 1990s and was finally applied to load forecasting [20]. However, it encountered significant challenges in other applications due to limitations with both software and hardware resources to process large input dimensions. The required time for model training and validation was hence prolonged. Another instance of the data-driven approach is expert systems in the 1980s to 1990s [21]. Dedicated research efforts were made during those times but ended prematurely due to the complexity of rules and the constant need to update the rules according to the changes in system conditions.

Thanks to the renaissance of AI/ML research and access to more powerful computer hardware today, the industry is ready to revisit the data-driven approaches to power systems and re-evaluate the possibility of bringing the power of AI and ML to power systems. The idea is strengthened in parallel with concurrent investment in communication and metering infrastructures that allow power utilities and grid operators to obtain high-speed and low-cost data. Some emerging data-driven use cases include behind the meter (BTM) solar generation prediction [22], anomaly detection [23], cyber-security-based intrusion detection systems [24], Locational Marginal Price (LMP) forecast and autonomous voltage control [25].

Before cloud became available, most AI/ML workloads were isolated and cost-prohibitive due to the algorithms' data storage and processing requirements. Still, the economics of the cloud now enable ML capabilities in a highly accessible manner. Moreover, they have unlocked the ability to constantly switch to evolving algorithms without the worry and effort to manage upgrades to hardware and licensing. By aggregating data in one place and addressing the needs for these data-driven approaches, the cloud is opening doors to more real-time analysis for grid operation.

*4) Wide-area Monitoring and Situational Awareness*

No nationwide monitoring infrastructure for the bulk power grid exists today. Each region, independent of its location, whether North America, Europe or Asia, is managed by a regional system operator. The regional system operator coordinates with sublevel local control centers to ensure reliable monitoring and control of the grid. These regional coordinators and transmission companies have their own software solutions for data acquisition, orchestration, and visualization. Data sharing occurs across regional "borders" to the extent necessary to accomplish basic coordination, but on a need-to-know basis as defined by narrowly scoped peering agreements [26]. As the bulk power grid evolves in response to changes in electricity generation and consumption patterns, broader system monitoring and control are required. Such wide-area monitoring and situation awareness are becoming increasingly vital in enabling early warnings of potential risks and expediting coordinated control and problem-solving.

Cloud technology offers an ideal environment for multilateral data and results exchange because of its massive network interconnectivity based on dedicated CSP networks and the public Internet. The downstream applications that consume such data or results can also be hosted on the cloud to provide a shared online solution independent of spatial or size constraints to all collaborators. Over the last decade of development, cloud technology has incorporated many features to support massive real-time data streaming, cross-account/cross-region data sharing and fast content delivery network, making cloud technology the most efficient and economical solution for wide-area grid monitoring. These features, available through a variety of services by CSPs, can provide consistent, secure and scalable services to the power system entities for data sharing and collaboration.

*5) Real-time (backup) Operations and Control*

Power grid operators are obliged to keep their essential businesses running even during those unusual times when the control center facilities are unavailable or inaccessible. For example, during the outbreak of COVID-19, the grid operators had to run the grid reliably and securely, even with an increased likelihood of operational disruption. During these times, operators lacked timely support from IT and operation engineers as they were required to work from home. Many essential applications for generation dispatch did not allow remote access [27], increasing the challenges faced by grid operators. A new and emerging challenge for system operators is increasing the fault tolerance of essential controls to keep the lights on during such circumstances.

With the aid of cloud computing technology, some of the Bulk Electric System (BES) reliability services can be securely backed up through the CSPs, which offer far more infrastructure scalability and resilience for individual power entities. Whenever the local control facilities, such as software applications, communication networks and physical servers, are unavailable or inaccessible, the backup control system on the cloud can be manually activated to take over crucial operation tasks to maintain business continuity. With proper solution architectures, cloud services can help users overcome various types of faults, including single or multiple server failures, data center outages or any application component fault. For instance, an auto-scaling policy set up in the cloud can help users to maintain a fixed number of servers even though some virtual machines fail to spin up or become unhealthy after being started [28]. Besides, component resilience can be achieved through container orchestration solutions like Kubernetes, i.e., the failed components will be automatically replaced with new ones to recover from errors gracefully without degrading the user experience [29]. Since the cloud-hosted control system can be accessed from anywhere using secure connections and validated identities through granular access control, the operators can continue performing their job functions more safely and reliably during abnormal times and avoid potential disruption to operational continuity.

Cloud technology to backup fundamental grid controls delivers a higher fault tolerance achieved via redundant and fault-tolerant architectures. Unlike traditional control center configurations (i.e., main/backup) that are built independently but are located geographically close, the cloud workload on which the control center's key functions are implemented can be hosted in various data centers across a larger geographical footprint, reducing the likelihood of a single fault affecting operations. Even if a catastrophic event were to impact multiple cloud data centers in a specific region, these functions could be quickly switched or redeployed to alternative data centers in a different region that are geographically distant from the affected one. Many cloud providers now offer their own Domain Name Server (DNS) service, making a regional failover quick and imperceptible to the users. A region-wide extreme weather event is an excellent example of this scenario: A hurricane or a snowstorm sweeps over the Northeastern US, putting two independent control centers at high risk of telecommunication or power supply disruption. In case of an IT or OT infrastructure disturbance, the backup control service hosted in the cloud would serve as the "spare tire." Since the cloud workload can be hosted in a different regions separate from Northeast, e.g., in a data center in Southern California, the grid operator can resume system operation and control capability with ease through a secure and encrypted tunnel to the cloud service irrespective of the access location.

*6) Operational Efficiency and Customer Experience Improvement*

A distribution utility is a link that connects consumers with the bulk grid regardless of the generation fuel source. Today's residential energy customers are asking for more from their electricity providers: more clean energy options, more information transparency, and higher Quality of Service (QoS). This new reality of increased consumer expectations and desire has led to electricity providers offering many new products and services beyond basic electricity delivery. For example, a user-friendly, online power outage map with an integrated incident reporting system is a fundamental need for today's customers. With the data collected from AMI meters, user feedback collected through the customer reporting system and the data from external sources such as weather forecasts and asset management systems, the utility companies can perform complex analytics to enhance the customer experience and operational efficiency. Undoubtedly, cloud computing technology will facilitate this process because it simplifies the setup of a centralized repository to accumulate and store data from heterogeneous sources, also known as "data lake" [30]. A data lake can include structured data from relational databases (rows and columns), semi-structured data such as CSV, JSON and more, unstructured data (documents, etc.) and binary data such as images or videos that consumers upload. In addition, cloud-based data lakes offer benefits such as auto-scaling and provide utility companies and their customers a unified entry to access their heterogeneous data in a centralized location. In contrast, traditional on-premises solutions often form silos of information across different systems and require additional integration efforts to use the data. Furthermore, cloud technology makes it easy to create, configure, and implement scalable mobile applications that seamlessly integrate them with backend systems where advanced analytics on heterogeneous data is performed.

*7) Advanced Distribution Management*

Power & Utilities is known as a legacy industry, with many devices, processes and computations that have not yet taken full advantage of the recent advances in power electronics and software engineering. Moreover, there is a growing number of IoT and/or power electronic devices and sensors in measurement and control points spread across the electric grid. Many of these new sensors and devices are installed in the distribution grid near the end-consumer, requiring a scalable, secure, reliable, and intelligent Advanced Distribution Management System (ADMS).

ADMS provides numerous benefits. It improves situational awareness by providing a single view of system operations near-real-time, ranging from distribution substations to customer premises and other utility systems. The cloud is an optimal location to host ADMS as it provides secure integration options for IoT devices and sensors and other sensors and utility systems. Cloud technology also supports highly scalable data ingestion patterns: from real-time data streaming to batch data ingestion from traditional systems.

By supporting various data ingestion patterns coupled with low-cost, durable storage solutions, the cloud enables data-driven innovation instead of capturing and storing only a subset of the data due to cost and computing resources constraints. The availability of more data from cloud technology allows all data to be analyzed in tandem with physics-based electric grid models to automate the identification of data anomalies, model inaccuracies and gain new insights into grid operation. The cloud-based ADMS can efficiently analyze models with millions of system components by using horizontal and vertical scaling.

Cloud-based ADMS can utilize serverless, extensible, and event-driven architectures to operate the system reliably during unplanned events (e.g., faults, fault-induced switching, weather changes, measurement/control failures). More specifically, the serverless architecture allows modular applications to be added and updated over time to support event-driven workflows in response to a rapidly evolving grid with increasing frequency of unexpected scenarios mentioned above. These features also position cloud-based ADMS as an attractive choice for future grid operation that incorporates new elements such as large-scale Plug-in Electric Vehicles or PEV integration.

*8) Distributed Energy Resources Management*

The modern electric grid is rapidly evolving with significant advancements in the low-voltage distribution end. In part, the evolution is driven by an explosion of agile, smart, Internet-connected, and low-cost DERs. The phenomenon is underscored in recent industry reports. A Wood Mackenzie Ltd. Report [31] estimates DERs capacity in the U.S. will reach 387 gigawatts by 2025. Another report from Australia [32] predicts that 40% of Australian energy customers will use DERs by 2027. These DERs are either low-power or control low-power equipment and individually do not amount to a significant impact on the grid. However, once aggregated, they can provide significant value in strengthening grid reliability or resiliency. DERs can provide grid services such as demand charge reduction, power factor correction, demand response and resiliency improvement. These unique facets of DERs have precipitated the development of new products [33] [34] from various entities that enable aggregation and monetization of DERs, many of which aim to use aggregated resources in the bulk energy market to optimize asset profitability.

Cloud-based technologies are an ideal choice for these products as they can unify control of various participating DERs in one central location under the authority of a designated market participant. Such a unified cloud-based environment eases the management of participating DERs through a synchronous collection of data from various controllable devices and facilitates large-scale analytics, which otherwise could not be performed in remote locations. The approach itself is more of a necessity than a choice due to the geographical spacing of the DER devices, which are typically located with low computing resources. Nonetheless, almost all of these resources are connected to the Internet through IoT-based sensors. They hence enable low-cost and low-effort cloud aggregation solutions via general API-based products [35].

In the DERs connected grid, a single gigawatt resource is replaced via hundreds of thousands of small DER resources. Optimal use of these DER resources, the DER aggregation and dispatch algorithms require advanced computing resources as the number of decision variables rises exponentially. In the future, cloud technology for DER aggregation and services is expected to expand significantly to accommodate new-age resources (such as EV batteries, home-battery systems, and smart thermostats) and new-age grid analytics. These will include state-of-the-art optimizations, artificial intelligence, and data mining algorithms. Once again, the cloud's elastic computing capabilities, which scale up and down in tandem with needs, will reduce or eliminate the need for hefty financial investment in large inelastic on-premises infrastructure, data center infrastructure and personnel.

*9) Market Settlement*

Electricity is continuously generated and consumed on a 24-hour clock, but settlement periods are defined as distinct time frames, e.g., an hour in North America and 30 minutes in Australia. With the ever-growing integration of renewable resources and energy storage, the owners of resources are expecting to bid (offer) more frequently so that they can respond quickly to electronic dispatch instructions to realize most of the market efficiency gains due to the aligned settlement of five-minute profiled MWh intervals and five-minute energy and reserve pricing. In short, reducing the settlement time blocks can better compensate market participants for the real-time energy and reserve products they are delivering. However, a shorter settlement interval places a higher requirement on the market operators' computational and data management capability. By exploiting elastic computing and big data infrastructure through the cloud, energy market operators can easily acquire the ability to process vast amounts of data promptly while significantly reducing operating expenses for themselves and the market participants.

*10) Collaborative System Modeling and Hybrid Simulation*

The cloud can be a central hub for collaborative model development and Transmission-and-Distribution (T&D) co-simulation. The recent issuance of FERC Order 2222 opens up a new opportunity for DERs to participate in the wholesale electricity market in North America, leading to an increase of joint T&D network studies following a hybrid co-simulation approach. These studies include but are not limited to: i) coupled T&D time-domain co-simulation combining electromagnetic transient (EMT) simulation with electromechanical dynamic simulation and ii) coupled T&D steady-state power flow and optimizations. In general, these combined T&D simulations can span up to hundreds of millions of solution variables with data scattered across many geographically separated locations. Solving analysis problems of this scale is not possible on a single compute node. While traditional on-premises computing clusters with limited resources (i.e., number of cores and available memory) for parallelism can conduct such hybrid simulations, their efficiency, in terms of speed and robustness, is low due to the need for high processing and memory requirements and high-volume data exchange between the various entities. With access to cloud scalability and cloud elasticity, the co-simulation of large T&D models can be fully accelerated. With the ability to scale the compute and memory resources at will, the industry can even incorporate other factors such as weather, finance and fuel constraints into modeling and simulation computing and storage resource limitations. One such cloud-based co-simulation approach has been validated by a consortium of U.S. national labs in a recent co-simulation platform HELICS [36].

Furthermore, cloud technology can also facilitate the collaboration of model development between the grid operators and distribution companies with its easy-to-access and easy-to-share characteristics, which is particularly helpful when the personnel is geographically separated. The COVID-19 pandemic has highlighted the heightened need for such collaborative effort. Moreover, since data security is sensitive to power grids, the cloud can be advantageous when running distributed simulations or optimizations, wherein data or models are spread across multiple utilities. Cloud allows various utilities to store and protect their data privately while exchanging minimum standard datasets with other entities to responsibly run the distributed simulation and optimization studies.

*11) Coordinated System Operation Drill*

Changes to electric grid operations due to the increased penetration of DERs, higher instances of extreme weather events, threats of cyber-physical attacks and loss of know-how due to the retiring workforce prompt updated training approaches for operators. Increasing drill participation by role-playing the different jobs can help operate the power grid more reliably during such events. However, such drills must be done in a simulated environment to avoid impacting the day-to-day operations of the grid. With access to a realistic simulation environment, multiple participants can collaboratively discuss and analyze extreme events. They can replay, rewind, or reload any prior or future scenario with no restriction on time or frequency of the analysis.

The advances in cloud computing allow the development of a simulation environment that enables real-time coordination among multiple participants, including reliability coordinators, balancing authorities, transmission operators, generator operators, and distribution operators. A cloud-based simulator can help bridge the gaps among planners, engineers, managers, system operators and cyber defenders. They can all observe and react to system behaviors such as exceedances of MW transfers limits, extreme voltage violations, and large operating angles using one real-time tool.

### III. CHALLENGES OF CLOUD ADOPTION IN POWER INDUSTRY

This section discusses the challenges of cloud adoption from the perspective of a) power grid users and b) software vendors. We provide a brief discussion on compliance-related challenges for cloud adoption as well. We provide corresponding guidance to overcome these challenges in the next section.

*A. Challenges for Power Grid Operators and Utilities*

The reasons that make average power utilities and grid operators hesitate to adopt cloud technology vary from organization to organization. Generally, there are no challenges from the technical perspective. Any on-premises solutions that have been realized technically can be redesigned and implemented in the cloud. Still, according to a survey conducted by our task force, the top three concerns for the power utilities and grid operators are cloud security, service reliability and cost.

*1) Cloud Security and Responsibilities in the Cloud*

Cloud security is a key concept to cloud adoption. Equally critical is the question of who is responsible for this security? As shown in Fig. 3, cloud security responsibilities are split between the power system users and the CSP like AWS and Azure. Such a responsibility division pattern is called Shared Responsibility Model (SRM) [37] [38]. Depending on which cloud service model is adopted, the users bear different areas of responsibility for cloud security.

For instance, in the Infrastructure-as-a-Service (IaaS) model, the CSP is responsible for the security of the underlying infrastructure that supports the cloud (security of the cloud). At the same time, users are responsible for anything they store on the cloud or connect to the cloud (security in the cloud). Even if a Software-as-a-Service (SaaS) solution is selected, users must consider data security in transit. In a nutshell, users should always be aware of their responsibility when using a public cloud service. Also, not all SaaS solutions are directly provided by CSPs. Independent Software Vendors (ISVs) offer many such solutions and host their service on a CSP's environment or an on-premises data center operated by the ISV. These differences are not always apparent to users, stressing the importance of confirming details and assigning responsibilities to protect their data and systems.

Guidance for cloud security is discussed in Section IV.A.2) of the paper.

*2) Service Reliability Related Challenges*

In addition to the security responsibility, service reliability is another concern for power utilities when using the cloud. Service reliability comes in many flavors, which are further explained below:

*a) Network Latency*

All communication over a network experiences latency, depending on the distance a network transmission must travel and the transmission speed of the network. Low latencies are desirable as high latencies can adversely affect the business services that depend on network communications, which might not meet the needs for fast transmission of information in some scenarios of power grid operators and utilities. For these users, using a cloud service over a network with high latency may result in erroneous decision-making, inaccurate control orders, or system malfunction. Such users in the power industry should be considerate of network latency when selecting both cloud providers and network connectivity options, to ensure that the network communication speed selected are adequate for their targeted cloud workloads.

*b) Network Connection Disruption*

Network connectivity, not necessarily Internet, is essential for a consumer to access cloud services. Power utilities choose between the public Internet and specialized telecommunication services that cloud providers offer in conjunction with telco companies (i.e., private networks). While private dedicated connections provide much higher reliability, they are susceptible to network outages like any other network. Architecture and failover options help handle potential network disruptions and alleviate the impact on grid workloads. Recovery Time Objective and different workload needs can be

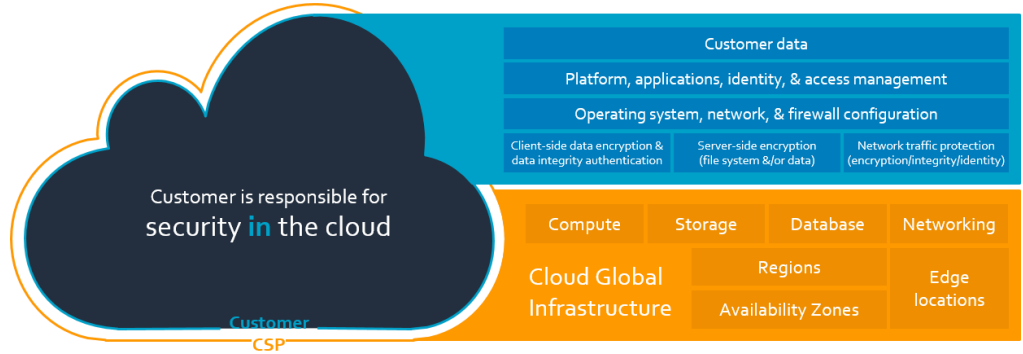

Fig. 3. Shared responsibility model (source:aws.amazon.com); The orange section represents the underlying infrastructure, which is the CSP's responsibility. The layer on top of the underlying infrastructure stands for the users' responsibility

considered in designing the cloud deployment.

*c) Cloud Service Outage*

Another outage that is even rarer than the network outage is the cloud service outage. Unlike the solutions customers bring on-premises to operate independently of vendor support, users on the public cloud depend on near-real-time provisioning of services by providers. Any duration of cloud service outage will lead to an interruption of the user's business process.

*3) Cost Challenges*

Another hurdle for utility companies to migrate their application portfolio to the cloud is the financial categorization of cloud expense. While the purchase of infrastructure such as in-house servers and their peripherals is considered as a capital investment and these can be recovered from utility's rate base [39], paying for the cloud services is currently regarded as an expense that falls into the category of Operation and Maintenance (O&M) cost. The O&M cost is not recoverable by the utility companies. Guidance in the form of solutions or mitigation measures for the above-stated concerns related to service reliability and cost of cloud application are provided in Section IV.A.3) and Section IV.A.4) respectively.

*B. Challenges for Software Vendors of Cloud Applications*

To practitioners in the electric energy sector, cloud computing is an emerging technology. Therefore software vendors at the frontier of providing cloud services are likely to face challenges as well. From a software vendor's perspective, these key challenges can be categorized into the following: the software design pattern, the licensing mechanism, and the pricing model.

*1) Software Design Challenges*

We consider the key design challenges for cloud adoption below:

*a) Monolithic architecture*

Most applications running in production systems today have monolithic architecture. Often, these can be ill-suited for cloud adoption, which has led to many companies realizing that simply moving their legacy system to the cloud either brings them marginal benefits or, even worse, negative benefits with unexpected problems. According to Google [40], challenges with monolithic architectures are four-fold: 1) fault isolation cannot be contained; 2) they are hard to scale; 3) deployments are cumbersome and time-consuming; 4) and they require a long-term commitment to a particular technology stack.

Therefore, for cloud-based software, there is a need to evaluate other software architectures. Section IV.B provides guidance on software architecture for the cloud.

*b) Software portability to cloud*

Software portability is another challenge with cloud infrastructure. Today, most grid software is compiled for a specific operating system and depends on routines for specific hardware and libraries. Hosting these tools on the cloud can be time-consuming and tedious as these would require the re-build of software based on the choice of the operating system, system libraries, and hardware. Therefore, software architects must consider advanced techniques (such as containerization, etc.) for easy portability to the cloud. Section IV.B.1)b) gives guidance on managing software portability for the cloud.

*c) Inefficient scalable and parallelizable code and algorithms*

Existing power systems software tools are generally not designed for high-performance computing and are often written to work on desktop computers using centralized memory, and single compute cores. Therefore, most of these tools will not take advantage of many cloud features when hosted "as is" in the cloud without modifications. For example, to succeed in high-performance computing applications, the software will have to adapt to easily parallelizable algorithms and operate on distributed memory resources. Section IV.B.1)c) provides guidance on adopting the software for enabling high-performance computing on the cloud.

Most software in the power industry today is stateful. In a nutshell, whether an application is stateless or stateful depends on where the "states", such as user profile information and client sessions, are stored. While stateful applications maintain the state data on the server itself, stateless applications put them on a remote platform. Stateful applications have superior performance because they don't need to process as much data in each client request compared to stateless programs, but being "stateful" has challenges when it comes to working on the cloud. First, it restricts the scalability of resources because application states are stored on the server itself. Replicating these states to newly launched servers in response to varying demand increases the processing overhead, thus lowering the application performance. Furthermore, the users of such an application need to continue sending requests to the same system that maintains their state data; otherwise, they will lose historical context. As a result, the users may experience delays and shutdowns as traffic rises to the degree that the server cannot handle. Second, it reduces the visibility of interactions since a monitor would have to know the complete state of the server. Therefore, stateless programs should be considered for cloud-based software. More guidance for this is provided in Section IV.B.1)d) of the paper.

*2) Software Licensing Mechanisms and Challenges*

Commercial software applications are licensed in several ways. The licensing methods differ in the management interface, process, allocation and availability. Like their peers in other business domains, the power system software vendors usually adopt all well-known methods to license their products. The most common licensing methods include dongle license, node-locked license or single-use license, Bring Your Own License (BYOL) and network license (also called floating or concurrent license)

To understand the licensing mechanisms and their suitability for cloud adoption, let us consider the software programs for transmission study as an example. Table II summarizes the licensing options available through various commercial power system software tools used for transmission operation and planning study. It is observed that most of them support migrating the user's software-dependent workloads to the cloud, either by offering a cloud-compatible licensing option or through a separate version of the product. As of when this report was written, PLEXOS and PSS/E (still in the test phase) offer a subscription-based license option to support cloud adoption. MARS and MAPS are allowed to run in the cloud if the license file is uploaded to the same host as well. In this case, the license validity time depends on the contract between GE and the customer. TARA also works well with cloud, but its license needs to be updated every so often because the cloud borrowed license is temporary. The other tools currently have limited support or lend no support for use in the cloud. For example, DSATools uses broadcast approach to search the working servers, while broadcast and multicast are typically not supported by major cloud providers. For the sake of this, DSA servers are not able to do horizontal scaling (scale-out, i.e., dynamically adding more virtual machines). They can only be scaled up by requesting more resources (CPU/RAM/DISK) on the individual virtual machine. Therefore, it can be concluded that while different software providers have many licensing mechanisms for usage in the cloud, few of them are designed for effortless cloud use. Guidance for licensing mechanisms that are most suitable for cloud is given in Section IV.B.2).

TABLE II
A COMPARISON OF LICENSING OPTIONS BY SEVERAL COMMON SOFTWARE PROGRAMS FOR TRANSMISSION STUDY
(Blank indicates "not sure" this specific licensing option is supported or not)

|  | Dongle License | Node-locked License | BYOL | Network License | Subscription Licensing | Suitable for Cloud |
|---|---|---|---|---|---|---|
| EnergyExemplar PLEXOS |  |  |  | √ | √ (PLEXOS Connect) | Yes |
| Siemens PTI PSS/E | √ |  |  | √ (CodeMeter) |  | Yes |
| PowerGem TARA |  |  | √ |  |  | short-term |
| GE MARS/MAPS |  |  | √ |  |  | Yes, duration depends |
| PowerTech DSATools |  |  | √ | √ |  | Only on single VM |
| Mathworks MATLAB |  |  | √ | √ | √ (MATLAB in the Cloud) | Yes |
| PowerWorld |  | √ |  |  |  | No |
| ABB GridView |  | √ |  | √ (cloud version) |  | Yes |
| GE PSLF |  |  | √ |  |  | short-term |

*3) Software Pricing Models and Challenges*

Software pricing models for a product vary from vendor to vendor, but at present, they can be categorized into two primary categories: pricing by license and pricing by subscription (usage). The former approach is what the majority of power system software vendors are providing. This pricing model comprises a one-time provisioning fee and a periodical maintenance fee in terms of payment cycle. The model can further include an additional cost for add-on features. Some vendors will even charge for concurrent use and/or off-campus use. No matter how the fees are broken down, these are high costs to the utility users.

This kind of traditional pricing model is one of the factors that hinders many utility companies from utilizing the cloud technology voluntarily. Using this model will significantly reduce software licensing costs if they move their business solutions built upon these tools to the cloud. In a nutshell, the pay-by-license pricing is unsuitable for the cloud environment. It lacks the flexibility to enable users to rapidly launch an experimental or non-production workload in the cloud, e.g., for agile development or to test a new workflow. For a utility that wants to run a production workload in the cloud, the most likely outcome today is that they end up paying two sets of licenses – one set for cloud use and another for on-premises use. Given a fixed budget range, it would be a waste of on-campus licenses if they made a pessimistic estimate for the cloud usage (buying too few cloud licenses). Alternatively, they would constrain the on-premises workflow if the cloud license was over-provisioned. Therefore, other pricing mechanisms must be considered for cloud software.

For guidance on how to make cloud licensing and pricing options better adapt to cloud environments for power grids users, see Section IV.B.2).

*C. Compliance Related Challenges*

In addition to technical and financial concerns, power utilities are hesitant about cloud adoption for regulated workloads due to a lack of clarity for compliance obligations associated with using cloud technology to run either the mission-critical workloads or the non-critical workloads.

Critical workloads are described as any service or functionality that supports bulk power systems' continuous, safe, and reliable operation. In North America, critical workloads are mandated to comply with a dedicated set of standards, known as NERC CIP [41]. At the time of writing, the currently effective NERC CIP requirements are silent on using cloud technology. Their device-centric structure raises questions on how to demonstrate compliance to the requirements. For example, there is a heavy emphasis on the current definition of physical assets within the Electronic Security Perimeter (e.g., the particular term "in those devices" referring to BES Cyber Assets. A BES Cyber Asset is defined as a cyber asset that, if rendered unavailable, degraded, or misused, would adversely impact one or more facilities and systems within 15 minutes of its required operation, misoperation, or non-operation. The key cloud concepts such as virtualization, logical isolation, and multi-tenancy are not provisioned in the standards or the current definitions.

Non-critical applications, such as those use BES Cyber System Information (BCSI), can be done in the cloud. NERC is on record stating that Entities have BCSI in the cloud and are doing so compliantly. Additional guidance on how to use the cloud securely, reliably and compliantly is in the early stages. In June 2020, NERC released a guideline on BCSI and cloud solutions, which is a start, but of limited scope as its primary focus on the encryption of BCSI. Implementation Guidance on cloud encryption and BCSI is pending NERC endorsement.

Addressing compliance challenges requires collaboration among multiple parties, including the regulatory bodies, utilities, CSPs, ISVs, third-party auditors. General guidance on how to deal with compliance for cloud applications is given in Section IV.C.

## IV. CLOUD ADOPTION GUIDANCE

*A. Guidance for Power Grid Operators and Utilities*

The five pillars of cloud adoption are operational excellence, reliability, security, performance and cost optimization [42]. Cloud solutions provide excellent technological capabilities and benefits when it comes to any of these columns. However, there are no general cloud requirements that apply to all use cases. Instead, the criteria for cloud offerings vary from case by case basis. For example, the Round Trip Time (RTT) of an Automatic Generation Control (AGC) control solution should not exceed 1 second in principle. At the same time, the requirement of this metric for state estimation and Security-constrained Economic Dispatch (SCED) can be less stringent because their execution cycles are in the timescales of minutes. The tolerance for the RTT for cloud-based post-event analyses, e.g., transmission needs assessment, can be as high as hours.

Using a dedicated network connection over fiber-optic cable yields more negligible communication latency than using a public internet connection over coaxial cable, but the cost of the former solution is generally much higher. Utilities should evaluate all the implications when considering adoption of cloud-based solutions. In other words, users typically consider the cloud when they want to pursue benefits from one or more pillars mentioned above. The decision to migrate to the cloud should only be made after due diligence from the entity and identification of specific requirements to meet business objectives. Whether an application should migrate to cloud mainly depends on its criticality and scope of impact. Usually, any non-production application can be deployed in the cloud to pursue the cloud advantages. In contrast, the mission-critical systems in the production environment, e.g., a system with feedback control in the loop, such as AGC, require careful consideration in design and testing before migrating to the cloud. For systems that comprise both on-premises parts and cloud-based modules, or have both in-house deployments as well as replicas in the cloud, normally require coordination between them via cloud-based integration [43].

*1) Choice of Cloud Type and Service Model*

What cloud type to choose? Which service model is the most suitable for a business user, particularly a utility company? The answers vary from case to case depending on the IT budget, business characteristics and value proposition.

*a) Selection of Cloud Resource Model*

There is no single cloud computing type that is applicable for all needs. Companies should analyze the advantages and disadvantages of each cloud type and align their goal with these characteristics before making the final decision.

The characteristics of these cloud types are compared in the table below. Users should choose the cloud type by aligning their goals with the characteristics below.

TABLE III
CHARACTERISTICS OF DIFFERENT CLOUD TYPES

| Cloud Type \ Characteristics | Private Cloud | Public Cloud | Hybrid Cloud |
|---|---|---|---|
| Control | High | Low | High |
| Flexibility | Low | High | High |
| Scalability | Low | High | Medium |
| Reliability | Low | High | Medium |
| Cost | High | Low | Medium |
| Maintenance effort | High | Low | Medium |
| On-prem workload migration effort | Low | High | Medium |

*b) Selection of Cloud Service Model*

As described in Section II.A, a few cloud service models exist, and they differ on how the responsibilities are shared between the service provider and the customer. The more responsibility the customer shifts to the service provider, the more convenience they will enjoy, but on the other hand, the less control and transparency they will have over the cloud offering. Corresponding to the introduction above, we only provide guidance for the best fit scenarios for the power industry: the IaaS, PaaS and SaaS models.

As the most "hands-off" model for cloud users, SaaS fortifies intra-company and inter-company collaboration through its easy-to-access and easy-to-share features. Furthermore, the service in SaaS mode can become functional in no time. All it takes is that you sign up for it.

IaaS allows the users to scale the resources up and down to build their cloud solutions based on their needs. IaaS will be the best fit for the scenarios when users decide what applications to run in the cloud and require porting licenses from on-site systems.

PaaS offers support for multiple programming languages, which a software development company can build applications for different projects. With PaaS, enterprise customers can benefit from having a consistent platform and unified procedure to work on, which will help integrate your team dispersed across various locations.

Table IV lists a few selected use cases that fit best into the SasS, IaaS and PaaS models. It further distinguishes the three business models in terms of business drivers and security (trust vs. control). While common business drivers, such as cost savings, agility, faster innovation, global reach, and elasticity are applicable to all three cloud service models, Table IV calls out the most relevant business driver for each.

TABLE IV
A LIST OF SCENARIOS THAT FIT THREE COMMON SERVICE MODELS

|  | SaaS | PaaS | IaaS |
|---|---|---|---|
| Example Use Cases | Modeling and Simulation, Asset Management, Collaborative operation | Big data analytics, Machine learning DevOps | Planning studies, Wide-area monitoring and situational awareness, Power outage map and incident reporting |
| Most Relevant Business Driver | Faster innovation – focus on business value | Agility – experiment frequently and quickly | More customization options |
| Trust vs. Control | More trust in provider | | More user control |

### 2) Managing Security

In general, the scale of public cloud service providers allows significantly more investment in security controls, policing and countermeasures that almost any large company could afford. This is even more pronounced for smaller power utilities. However, security is a shared responsibility, and cloud users still have to invest in security controls applicable to their roles. Based on the Shared Responsibility Model (SRM) shown in Fig. 3, users are still responsible for "security in the cloud", which means controls to secure anything that the user puts in the cloud or connects to the cloud. There is no "hands-free" mode when it comes to cyber security controls in cloud adoption. Power utilities must ensure that professionals are in place to deploy the right security policy for their cloud-hosted services. Fig. 4 shows a comprehensive scheme to secure the cloud workloads. It depicts an organization of security control in the cloud and groups the security control domains at a high level. Again, this is a shared responsibility, so cloud users, ISVs and the CSPs will share responsibility for the controls, depending on which cloud service model the user adopts. While users may not be directly responsible for security control, they are still strongly encouraged to ensure that the vendors enforce the controls shown in the diagram. We further discuss the critical controls in more depth in the following paragraphs.

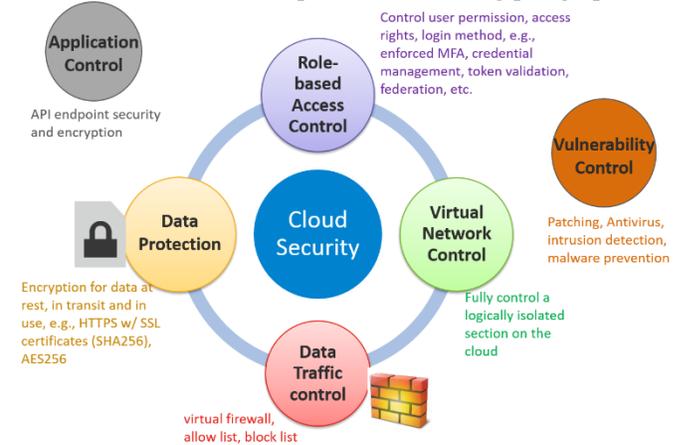

Fig. 4. Comprehensive security control in the cloud

*a) Virtual Network Control*

A Virtual Network (VNet) or Virtual Private Cloud (VPC) is a network environment dedicated to the user's account in the public cloud. This virtual network closely resembles a traditional network that one can operate in their own data center, with the benefits of using the scalable infrastructure. A VNet in the context of Azure or a VPC in AWS and Google Cloud logically isolates infrastructure by workload or organizational entity. The users can launch the cloud resources to meet any business needs and complete control in the virtual network. They can also create multiple subnets that span across AZs to define different access rules. Sometimes the users want to extend their network into the cloud using CSP's infrastructure without exposing the corporate network to the Internet. In this scenario, the workloads can be deployed in a private subnet of the virtual network, and a site-to-site IPSec VPN tunnel [44] can be established from their network to the cloud directly. Since there is no Internet gateway to enable communication over the Internet, the likelihood of cyber-attacks and data breaches is significantly reduced. The user should generally refrain from exposing any of their servers (including public-facing workloads) on the cloud to the Internet directly through a public subnet. Instead, consider using another layer of security between the server and the Internet.

*b) Identity and Access Management*

Identity and Access Management or IAM is critical to protect sensitive enterprise systems, assets and information from unauthorized access or use. The best practices for IAM on the cloud include but are not limited to:

<u>Use a strong password for account-level access</u> – CSPs highly recommend a strong password to help protect the account-level access to the cloud services console. Account administrators can define custom password policies to meet their organization's requirements. Typically, strong passwords should consist of uppercase and lowercase letters, numbers and special symbols, such as punctuation. The minimum length of the password should be 8 characters or more.

Enable MFA – For extra security, MFA should be enforced for all users in the account. With MFA, users have a device that generates a response to an authentication challenge. Both the user's credentials and the device-generated response are required to complete the sign-in process. If a user's password or access keys are compromised, the account resources are still secure due to the additional authentication requirement.

Create individual users for anyone who needs access to the cloud service – the root user's credentials and access key for programmatic requests should be locked away. Instead, create individual users for anyone who needs access to the account, including the system administrators. Give each IAM user zero permission at first while their account is being created; grant them only necessary permissions to fulfill their job duties by request.

Use roles to delegate permissions – Do not share credentials between accounts to allow users from one account to access resources in another account. Instead, designate the users in different accounts to assume different IAM roles for their access. Also, use role-based access for applications to access cloud resources. Unlike IAM users, roles are temporary credentials generated randomly and automatically rotated for whoever assumes the designated role.

Rotate credentials regularly – Enforce all users in the account to change the passwords and access keys regularly. That way, a compromised password or access key that is used to access the resources in the account with permission can be limited.

Integrate with existing identity providers – Most CSPs can integrate their IAM with popular identity providers to allow identity federation. By doing so, power utilities can centralize access controls, improve efficiency, and maintain compliance with processes and procedures.

*c) Vulnerability Control*

While utilities pursue rigorous vulnerability management in their on-premises systems, they may misplace their trust in cloud providers regarding vulnerability control in the cloud environment. However, the cloud providers are only responsible for securing the underlying infrastructure (the hardware and firmware). The customers must detect and address a vulnerability in their solutions on their own or through a trustworthy third party. The third-party cloud includes a vendor who delivers the solution via the cloud or a provider partner who offers professional vulnerability scanning and patch management service.

Gaining visibility into vulnerabilities in the code is key to reducing the attack surface and eliminating risk. Power utilities usually rely on security tools to evaluate the potential risks or deviations from the best practices in their applications. They should follow the same practices for cloud-based solutions. There are quite a few security assessment services for different cloud systems on the market, and utilities should use them. These services, updated regularly by security researchers and practitioners, can help improve the security and compliance of applications deployed in the cloud environment. It is more efficient and convenient for power system companies to adopt an IaaS-based solution to take advantage of such services. Management of the vulnerability can also be done by transferring the responsibilities to a reliable vendor, especially in the case of PaaS and SaaS service models. Finally, it should be noted that it is the utilities' responsibility to perform a regular review of what vulnerability controls have been adopted by the vendors and ensure they have minimum practices in place.

*d) Data Protection*

In addition to the IAM setups that help protect data, e.g., enforced MFA, user account separation from the root account, the following approaches should be considered as well by the utilities to protect the data further when it is: i) at rest, ii) in transit or iii) in use.

Data at rest refers to stored or archived information on some media and is not actively moving across devices or networks. Although data at rest is sometimes considered less vulnerable than data in transit, power system users should never leave them unprotected in any storage (cloud or on-prem). Protection of data at rest aims to secure inactive data, which remains in its state. Usually, encryption plays a significant role in data protection. It is a popular tool for securing data at rest since it prevents data visibility in the event of unauthorized access or theft. The data owners are obligated to know what encryption algorithms a cloud provider supports, their respective strengths or cracking difficulties, and what key management schemes they provide. The recommended encryption method to secure data at rest is Advanced Encryption Standard (AES) with a key length of 256 bit, i.e., AES-256. AES-256 is the most robust encryption standard that is commercially available today. It is practically unbreakable by brute force based on current computing power.

Using CSP-provided Key Management Services (KMS) is another way to protect data. Such services integrate with other CSP services to encrypt data without actual movement of key material. A user can simply specify the ID of the key to be used. All actions using such keys are logged to record by whom and when a key has been used. KMS systems also help with automated key rotation.

Likewise, data protection in transit is the protection of data when it is being moved from one location to another, particularly in this context, when it is being transferred from a local storage device to a cloud storage device or the other way round. Wherever data is moving, effective data protection measures for in-transit data is critical as data is often considered less secure while in motion. It is widespread for enterprises to encrypt sensitive data before moving and/or using encrypted connections such as HTTPS and SFTP (secured by security certificates via SSL or its successor TLS) to protect data contents in transit.

Data in use refers to data in computer memory. It is usually considered the cloud provider's responsibility to ensure the underlying hardware and OS are malware-free, based on the SRM. However, it is still the user's responsibility to confirm that the software packages, the application dependencies and container images that are deployed on the cloud are free of malicious code. This is similar to the job function of the IT department in utilities today, wherein they sanitize the software that is used locally.

*3) Service Reliability: Building Fault-tolerant Solutions*

Power system users should aim at building reliable, fault-tolerant, and highly available systems when considering cloud-native adoption or migrating on-premises workloads to the cloud. A fault-tolerant architecture helps the user to ensure high availability and operational continuity in the event of failures of some components.

*a) Mitigation of Network Latency Impact*

The following techniques can be used to alleviate the impact of network delay on the business workloads:

Utilize CSP-provided dedicated networking: Many CSPs work with telecommunication providers to offer dedicated bandwidth private fiber connections to the CSPs region from the customer's location. This allows customers to have known stable performance with no risk of network latency due to general Internet traffic. AWS Direct Connect and Azure ExpressRoute are examples of services where a dedicated fiber connection is established between a utility's data center and the cloud.

Shift data processing to the edge: For data collected by IoT devices, they can be processed at the edge, which is known as edge computing, to further address the concern around latency by reducing the payload for transmission.

Leverage load-balancing: The users can distribute traffic across multiple resources or services employing load balancing to allow the workload to use the cloud's elasticity maximally.

Choose workload's location based on network requirements: Use appropriate cloud locations to reduce network latency or improve throughput. The users can select those edge locations which have low delays to host the workload. When the edge location becomes congested, use the local balancer to reroute the traffic to other edge locations where the latency is low. Distribute workload across multiple locations if needed.

Optimize network configuration based on metrics: Use collected and analyzed data to make informed decisions about optimizing your network configuration. Measure the impact of those changes and use the impact measurements to make future decisions. For example, if the network bandwidth is low, it might serve well to increase the bandwidth. In addition, consider options for higher quality bandwidth with less packet loss and provision for retransmission. Usually, business-class Internet services and dedicated circuit networking can meet these goals.

Critical workloads, especially those supporting real-time decision-making and bulk power system operation, are more sensitive to network latency. Therefore, apart from the general network optimization approaches mentioned above, additional countermeasures should be adopted *if there is a likelihood of hosting critical workloads on the cloud*. For instance, an offline work mode should be provisioned in the design of a cloud-based solution to allow the users to temporarily work on their problems using manually fed data (collected via secure phone calls by operators). Utilities may also consider CSP-provided hardware for such operations.

*b) Mitigation of Impact of Network Connection Disruption*

As long as the service is hosted in the cloud, theoretically, a user can access it from anywhere given access to an Internet connection (secured tunnel and identity authentication are usually required by utilities) or a dedicated network connection. It is recommended to have an independent connection at each of the multiple locations to achieve high resiliency. For example, grid operators typically have two independent control centers – the primary and the backup. As shown in Fig. 5 (a), each data center can connect to the cloud through independent telecommunication providers. Since an outage of one telecommunication provider usually does not occur concurrently with an outage of another provider, this provides inherent resiliency during ISP outages. Switching the connection from the down provider to the healthy one via another control center is a general fault-tolerant solution to counteract the impact of a network outage.

Enhanced resiliency can be achieved by separate connections terminating on different devices at more than one location. As illustrated by Fig. 5 (b), duplicate links are established to provide resilience against device failure, connectivity failure, and complete location failure, even for the same corporate data center.

Besides relying on the wired network to connect with the cloud, utilities can also employ advanced wireless communication technology such as 4G LTE-A and 5G to connect edge devices directly to the cloud. As the IoT and edge computing technology mature, the data collected by the edge devices can be locally processed, encrypted and then transmitted to the cloud to feed the hosted services without using utility data centers as the data hub or relay. The impact of wired network disruption can thus be mitigated by such "edge to cloud" connections, as shown in Fig. 5 (c).

It is wise to avoid putting all eggs in one basket for those mission-critical applications or functionalities as a modern society cannot bear sustained power outages. AGC and SCED represent two such examples of critical workloads that continuously maintain the balance between power supply and demand. If these were to move to the cloud in the future driven by key benefits (even though there is no such a need at this time), we must back up the process in local non-remote infrastructure, even when independent and redundant network connections are established for cloud workloads.

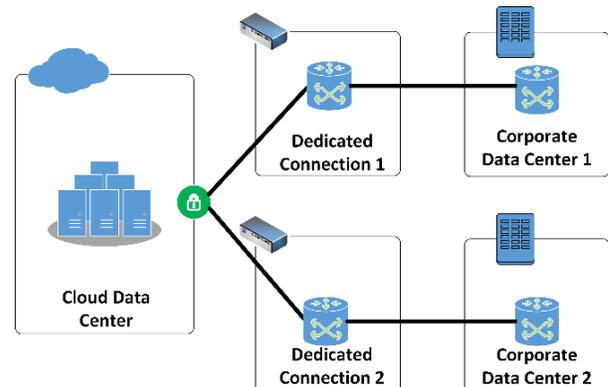

(a) Redundant connection from utility network to the cloud

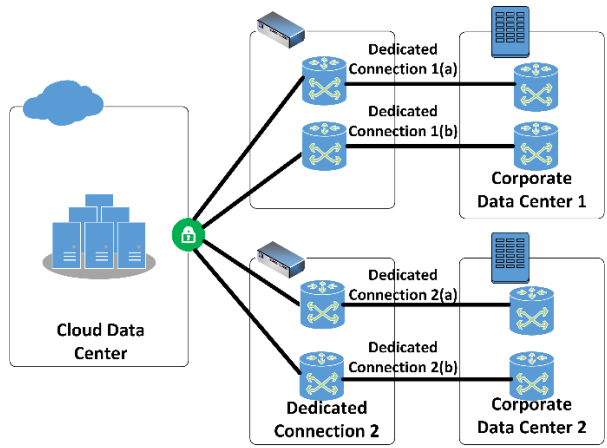

(b) Enhanced redundant connection from utility network to the cloud

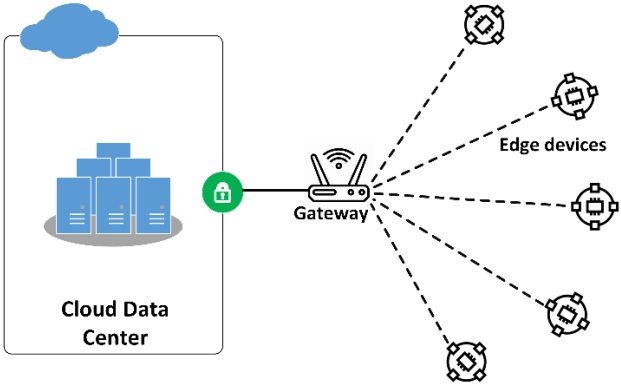

(c) Edge to the cloud connection

Fig. 5. Increase service fault tolerance through network redundancy

*c) Mitigation of Cloud Service Outage Impact*

To build fault-tolerant applications in the cloud, the users (or the solution vendor) need to first understand the cloud provider's global infrastructure. For example, users should grasp the concept of Region and (Availability) Zone (AZ) that major CSPs offer. High availability of solutions can be achieved by deploying the applications to span across multiple AZs or more than one cloud region. Placing various, redundant instances in distinct availability zones for each application tier (web interface, application backend, and database) creates a multi-site solution.

To minimize the business disruption or operational discontinuity, the utilities going to the cloud are recommended to adopt a redundant and fault-tolerant architecture that allows them to quickly switch their workloads to a different AZ in case of a single-location outage. Meanwhile, they need to consider duplicating the infrastructure of the running workloads in another region so that desired resources and dependencies can be spun up rapidly to continue the service when there is a region-wide outage. In addition, the users should also make their disaster recovery plans, e.g., identifying critical data and performing cross-region backups for them, to ensure the business continuity in the face of a man-made or natural disaster. As shown in Fig. 6, utilities can increase the fault tolerance of their cloud workloads and gain disaster recovery capability through backups in multiple AZs and Regions.

*4) Guidance on Cost Optimization*

Utilities today are hesitant about cloud adoption due to budget concerns, but the reality is that they can take control of cost and continuously optimize it on the cloud. With a meticulous analysis of their demand profiles, the utilities can take a few steps to quickly develop a plan that meets their financial needs while building secure, scalable and fault-tolerant solutions for their business needs [45] [46].

The first step is to choose a suitable pricing model. CSPs usually price the resources that users request based on the capacity inventory and the term of service. For example, in the context of an elastic computing service, users may see "Spot Instance (Virtual Machine)," "On-demand Instance (Virtual Machine) and "Reserved Instance (Virtual Machine)." While "On-demand" normally meets users' needs, "Reserved" pricing provides a remarkable cost reduction on top of "On-demand" if the target workload is consistent over a period of time and users would like to sign a long-term agreement (1 year ~ 3 years) with the CSP. And "Spot" pricing, which works like an auction/bidding process for CSP's surplus capacity after meeting the "On-demand" and "Reserved" needs, offers a further discount for the users to run stateless, fault-tolerant and flexible applications such as big data and planning studies.

The next step is to match capacity with demand. Cost can be optimized when resources are aptly sized. To line up the capacity with the real-time demand, the utilities or their designated solution vendors should consider an auto-scaling strategy to allocate the resources to match the performance needs dynamically.

Moreover, users may want to implement processes to identify wastage of resources since it will further help them optimize their cost by shutting down unused or under-utilized resources. Various CSPs offer services to help their customers identify these idle resources, such as AWS Trusted Advisor and Azure Advisor.

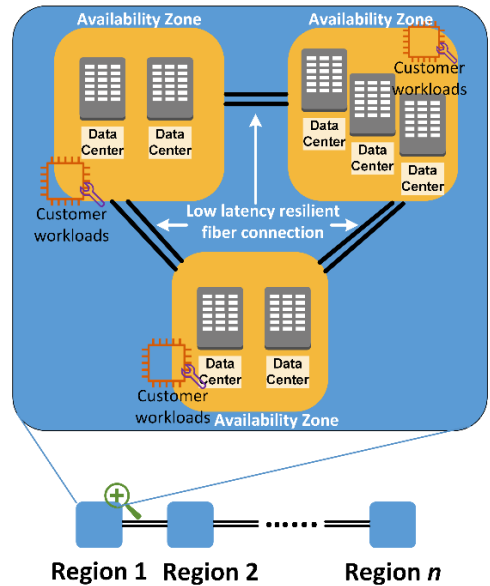

Fig. 6. Increase service fault tolerance and disaster recovery capability through backups in multiple AZs and Regions

### B. Guidance for Software Vendors

*1) Software Design to Support Cloud Adoption*

The grid software must be architected and designed to be cloud-friendly to facilitate cloud adoption by the utilities. Power system software vendors and in-house developers should consider the following while designing and developing cloud-based solutions:

*a) Microservices instead of monolithic architecture*

Due to challenges with monolithic architectures for cloud infrastructures, microservices architecture has emerged as a prime candidate for maximizing the benefits of moving systems to the cloud. It brings design principles to make software components loosely coupled and service-oriented. A fundamental principle of microservices is that each service manages its data. Two services should not share data storage. Instead, each service is responsible for its private data store [47], i.e., Binary Large Object (BLOB), as shown in Fig. 7. The modularity introduced by this architecture also offers ease and agility to keep up with the accelerated pace of business. Due to module independence, any service can be scaled and deployed separately in a microservices-based application. Besides, the modular characteristic of microservices inherently enhances security and fault isolation.

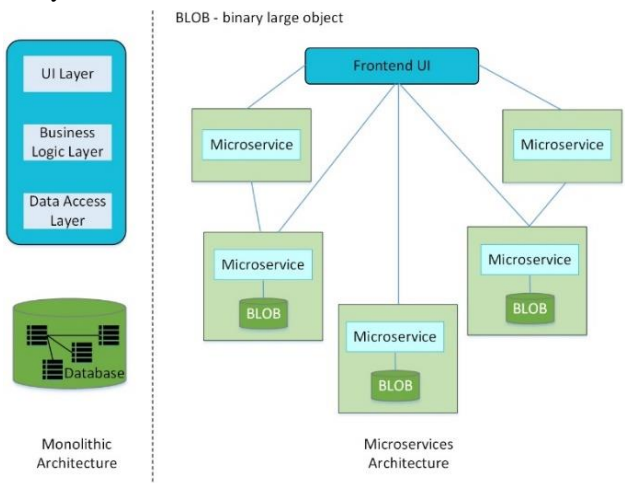

Fig. 7. Comparison of monolithic architecture and microservices architecture

Although microservices architecture has many advantages, it is not an omnipotent solution for cloud-native applications ("Cloud-native" is an approach to building and running applications that exploits the benefits of the cloud computing delivery model. Applications that are built in the "cloud-native" style are a collection of small, independent, and loosely coupled services, in contrast to the non-cloud-native programs, which are usually designed in a monolithic fashion) [40]. While bringing agility and speed to software development, microservices architecture sacrifices the operational efficiency as there are many more parts/modules than the monolithic approach. The software vendors should avoid the overuse of the microservices layout for the development of cloud-native applications. For instance, a power flow analyzer or state estimator should not adopt a microservice architecture because both processes are tightly coupled. Both programs can frequently invoke other methods such as network topology processors. Due to this, related processes are more likely to have a more considerable computation overhead when using a microservices architecture. However, a time-domain simulator for transient stability assessment can be orchestrated using microservices architecture. These analyses are independent processes given the same initial condition and can be launched in parallel.

*b) Use of containerization to improve software portability*

Containerization is a critical technology that vendors may want to consider for cloud-native application development. It significantly streamlines the deployment process because the development, test, and deployment environments are made consistent in containers to remove barriers between them. Containers allow the developers to package an application and its dependencies into one succinct manifest that can be easily ported to a virtual machine or cloud environment without rewriting the code.

Furthermore, containerization enables developers to scale the applications by changing specific components or service modules while keeping the remainders of software unchanged. For instance, in a container-based design, programmers can scale databases for network data or dynamic data and corresponding modules to support the increased processing load without scaling web server resources when the size of study cases varies. Employing some container orchestration tools like Kubernetes (k8s), the services can be scaled up or down automatically based on resource utilization, which aligns with the characteristics of cloud-native applications.

Container technology and microservices architecture are tightly coupled and usually adopted in cloud-oriented applications together since containerization provides distinct and natural boundaries between different microservices. However, using containers does not imply that the software follows the microservices architecture. Monolithic applications can be containerized too. Although both techniques are consistently utilized in application modernization, vendors should carefully consider them for development based on the software characteristics. The final goal of software portability is not hindered [48].

*c) Enabling HPC Capabilities*

The software vendors, particularly those who provide solutions for power system modeling, analysis and simulation, should accommodate critical technologies that enable high-performance computing when designing their software. More succinctly, to unleash the power of HPC infrastructure, they must employ expertise in parallel computing algorithms design and programming. From the algorithm design perspective, the vendors need to develop new domain-specific parallel computing algorithms for compute-intensive tasks and memory-intensive tasks, for example, large-scale optimal power flow, unit commitment, transient stability, and electromechanical simulation. They should also apply powerful parallel data processing methods in their applications when there are massive data ingestion streams. Besides, new data storage techniques must also be considered for the seamless transfer of data between storage bins and compute units.

Although the relational database is still the mainstream in the power industry, other types of databases such as NoSQL database (key-value store, document store, graph, etc.) and distributed databases are beginning to play a more significant role as the data yield capacity and locations in the grid where the data is collected are growing. From the programming point of view, new architectures should be adopted in the software design to support multi-thread and multi-core processing. For instance, one crucial factor that significantly impacts parallel computing performance is the coordination between submodules. Methods to efficiently distribute the software functionality across these computing resources are expected. If these considerations are not adequately addressed during design, applications meant to be run in multiprocessor, multi-core environments could end up with severe and hard-to-find performance issues [49]. Finally, for an effective HPC software design, users should be given the flexibility to opt-in/out of the parallel processing and the choice to choose the computing hardware (e.g., no of cores, RAM usage etc.).

    *d) Use of stateless protocol*

A stateless protocol should be considered to make an application compatible with microservices architecture, especially in the cloud. In general, moving to be "stateless" protocol brings several significant benefits, including 1) removing the overhead to create/use sessions; 2) providing resiliency against system failures and recovery strategies in the event of failures; 3) allowing consistency across various applications; 4) scaling processing capacity up and down to handle variances in traffic.

Determining the best protocol for cloud-native systems depends on the business model and the use case. For example, while a stateless application is ideal for short circuit fault analysis, a stateful design will be more suitable for forecasting system load or power flow on the tie-lines in the next few minutes/hours because of the historical context. At present, however, power system utilities overuse stateful protocol and lose on advantages of stateless protocol when using the cloud. In conclusion, a software vendor should neither be over-dependent on the traditional stateful design nor abuse stateless models.

The guidance given above is general recommendations mainly for designing non-mission-critical software to better support cloud adoption. Risk-based cascading analysis [50] is an excellent example of how incorporating this guidance in the design will help an application leverage the cloud advantages. To predict the probability of transmission outages at different locations under extreme weather conditions and subsequently perform contingency analysis based on the predictions, the application needs to include several core components: 1) weather module, which continuously fetches weather forecast data; 2) transmission structure module, which models transmission towers and conductors as well as their fragility curves; 3) geography module which returns the elevation data; 4) vegetation system model which acquires the tree types, tree height and land cover data; and 5) contingency module which manages the contingencies to be studied under high-probability outages. Since these modules are loosely coupled but functionally independent, they are a good fit for microservices architecture and container technology to isolate any fault in the component. Making each module "stateless" is also beneficial because they don't rely on session-related information to generate output. Building them in a "stateless" fashion makes it easier to scale out to handle increasing demand. Lastly, carrying the HPC capability undoubtedly expedites the contingency analysis process.

For mission-critical applications and systems, such as SCADA and automatic generation control, guaranteeing bulk power system operation continuity, a careful study on whether they should be moved to the cloud is necessary. After the study, one should carefully evaluate whether any guidance in the paper would degrade their performance. The entity should closely examine any computation and communication overhead due to the adoption of guidances above.

  *2) Software Licensing and Pricing Suitable for Cloud*

The pricing models for cloud-native applications need to cooperate with proper licensing models. In addition to the network license model and BYOL model, the vendors should also consider licensing the software based on subscription. The subscription-based licensing model, together with the pay-by-subscription or pay-per-use pricing modes, are designed to be more cost-effective and convenient in the cloud. In the pay-by-subscription pricing method, users only need to pay for the use of software periodically based on the agreement between them and the vendor. The billing cycle can be daily, monthly, quarterly or annually. Regardless of how the software is utilized, the subscription fee is usually fixed, varying a little bit depending on the selected payment option. The second approach, pay-per-use, or rather, pay-as-you-go in the context of cloud computing, provides users a finer granularity of control overspending on the software service. The vendor only bills the users based on their usage (or time) in the cloud. From users' point of view, they no longer need to worry about whether the cloud license will be overprovisioned or under-provisioned because they only pay for the time when the service is being used.

Unfortunately, the subscription-based licensing and the two discussed pricing methods that rely on users' subscriptions have so far not been widely implemented by power system software vendors. The utility companies might be tolerable with the old-fashioned licensing and pricing options in the traditional IT environment. Still, they would expect a more flexible way for software licensing and charges when it comes to the cloud.

*C. Guidance on Security Compliance for Cloud Applications*

Compliance encompasses checks to confirm that a set of specified requirements are implemented and are operating as expected. Such conditions can be set internally within the organization to meet business, operational, or security objectives. They can also be required by regulatory standards such as the NERC CIP Standards.

  *1) Compliance Certifications and Authorization*

Numerous well-established security assurance programs exist that standardize security assessment and authorization for cloud products and services, such as FedRAMP [51], which the U.S. federal agencies enforce. Such security assurance programs such as SOC, ISO, and FedRAMP, among others, require continuous monitoring and audit by cloud security experts. Third-party assurance reports support compliance demonstration for security "of" the cloud. Such certifications provide security assurance to cloud users as well as support achieving certifications of their own. For example, entities seeking SOC-2 certification can inherit a CSP's SOC-2 certification and focus their compliance efforts in fulfilling the requirements applicable to their "in" the cloud applications. Utilities should always research and investigate a cloud provider before using its services. One approach to identifying reliable cloud providers is to look at the certifications, accreditations, and regulatory controls that a cloud provider has earned or demonstrates.

Fig. 8 shows how users, ISVs and CSPs collaborate on assuring the security compliance authorization.

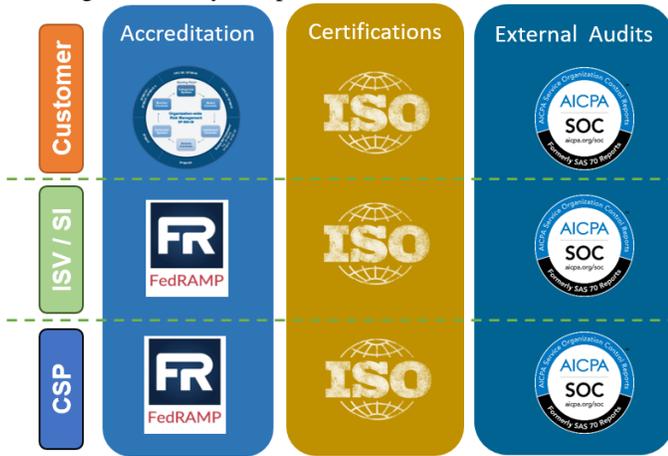

Fig. 8. Shared security assurance

In the model as shown in Fig. 8, customer is responsible for obtaining and maintaining certifications and accreditations through internal/external audits (as required) for their end-user system leveraging CSPs and vendors, while the vendor is responsible for obtaining and maintaining certifications and accreditations through external audits (as required) for their cloud offerings. For CSPs, they need to take responsibility for obtaining and maintaining infrastructure certifications and accreditations through external audits

*2) Critical Workload and CIP Compliance*

For power system users in North America subject to the CIP standards, Table V shows a mapping of NERC CIP standards to CSP security categories. These categories organize the essential capabilities to drive a cloud user's security culture. It also helps users to structure the selection of security controls to meet security and compliance needs.

TABLE V
A MAPPING OF NERC CIP STANDARDS TO CSP SECURITY CATEGORIES

| NERC Standard | NERC Standard Description | Corresponding CSP Security Categories |
|---|---|---|
| CIP-002 | BES Cyber System Categorization | Customer Determination* |
| CIP-003 | Security Management Controls | Governance |
| CIP-004 | Personnel & Training | IAM |
| CIP-005 | Electronic Security Perimeters | IAM<br>Infrastructure Protection |
| CIP-006 | Physical Security of BES Cyber Systems | Infrastructure Protection |
| CIP-007 | Systems Security Management | Detection<br>Infrastructure Protection |
| CIP-008 | Incident Reporting and Response Planning | Incident Response |
| CIP-009 | Recovery Plans for BES Cyber Systems | Incident Response |
| CIP-010 | Configuration Change Management and Vulnerability Assessments | Infrastructure Protection |
| CIP-011 | Information Protection | Data Protection |
| CIP-012 | Communications Networks | Data Protection |
| CIP-013 | Supply Chain Risk Management | Data Protection Detection<br>IAM<br>Incident Response<br>Infrastructure Protection |

* This is not a CSP security category. Power system companies must determine the risk and compliance category of a specific system.

Cloud solutions can meet the security objectives in CIP standards by implementing these security controls. CSPs have well-documented user guides, technical papers, guidance documentation, and interactive engagement that support their customers in implementing security controls and achieving a high level of security.

Security assurance certifications and authorizations present an opportunity for use in NERC CIP. A comparison between the FedRAMP Moderate control set and NERC CIP requirements reveals that the FedRAMP Moderate control baseline encompasses all NERC CIP requirements [52]. NERC's acceptance of FedRAMP as CIP compliant for security "of" the cloud can provide a streamlined compliance approach for power utilities and support efficient audit assessment while maintaining stringent security obligations. Registered Entities building on FedRAMP authorized infrastructure and services could focus their security controls and compliance assessment on security "in" the cloud.

*3) Non-critical Workload Compliance*

Non-critical workloads refer to applications or services that are neither mission-critical nor high-impact. In North America, such workloads are out of scope for the NERC CIP standards, but registered entities are still required to protect any Critical Energy Infrastructure Information (CEII) data involved in cloud adoption according to section 215A(d) of the Federal Power Act. The best approach to protect confidential data and user privacy is to follow the CSP-endorsed best security practices.

*4) Internal Compliance*

Power industry users often establish security expectations and internal compliance obligations to align and confirm enterprise-wide objectives. Security Frameworks such as the NIST Cyber Security Framework (CSF) [53] provide a holistic approach to security controls across networks. Entities can choose to meet compliance requirements from various security frameworks and/or regulatory standards and map them to their implemented controls.

## V. Use cases for Cloud Technology in Power Industry

This section summarizes a collection of real-world use cases for cloud technology in the power industry. The summary is provided in Table VI. The table breaks down each real-world use case in terms of the company, application area, cloud type and service model, upfront and operating cost and maintenance effort. It also summarizes the security control scheme and features benefits for each real-world use case. The use cases cover a wide variety of power system organizations (such as ISOs, utilities, and private vendors) as well as application areas (such as planning, operation and market related). The table also documents an optimal choice of cloud service model for each real-world use case. Finally, the table documents the benefits of cloud technology for each use case as drawn from specific business needs, drawn from Section II.C. We hope that these use cases can guide other cloud adoptees in the power sector. From these success stories, the reader can learn how the primary concerns over the cloud can be addressed or how the risk while migrating to cloud technology can be mitigated. As these use cases are collected through a participant survey, the details represent individual participants' responses and may vary from use-case to use-case. A summary of the use cases is given in Table VI and a more comprehensive discussion follows with various use cases categorized based on their utility.

TABLE VI
A Summary of Real-world Use Cases in the Power Industry
(The table groups similar cases in the same row. The superscripts indicate the case numbers, which link the security schemes to corresponding cases)

| Case No. | Company | Application Area | Reference Guidance | Cloud Type | Cloud Service Model | Upfront Cost | Operating Cost | Maintenance Effort | Security Control Scheme | Featured Benefits |
|---|---|---|---|---|---|---|---|---|---|---|
| 1, 2 | NYISO[1] ISO-NE[2] | Planning studies (Ref. [18] [54]) | IV.A. IV.B.1).c) IV.B.2) IV.C.3) IV.C.4) | Public (AWS) | IaaS | No | Cloud resources usage; software license fee | Low | IAM[1,2], data encryption[1,2], security group[1,2], enforced MFA[2], password rotation[2], role-based access[2], HTTPS/TLS[2] | Lower cost; better scalability, much less task completion time and a job waiting time |
| 3 | ISO-NE | Load forecasting | IV.A. IV.C.3) IV.C.4) | Public (AWS) | PaaS | No | Cloud resources usage | Low | IAM Role, data encryption, enforced MFA, API activity logging, HTTPS/TLS | Integrated process for data preparation, model building, training/tuning and testing to accelerate ML development |
| 4 | ISO-NE/ NYPA | Wide-area monitoring and data sharing (Ref. [55]) | IV.A. IV.C.3) IV.C.4) | Public (AWS) | IaaS | No | Cloud resources usage | Medium | VPN over SSH, AES256 data encryption; separate subnets with NAT instance | Lower cost; data sharing with ease, high fault tolerance, fast data recovery consistent display to enable real-time collaboration |
| 5 | ISO-NE | Backup control and emergency dispatch (Ref. [27]) | IV.A. IV.B.1).a) IV.B.1).d) IV.C.2) IV.C.4) | Public (AWS) | FaaS | No | Extremely low considering the event is rare | Low | HTTPS/TLS for data in transit, SSE-KMS for data at rest, access, IAM User/Role, API key, signed URL | Provide a backup solution to ensure operational continuity in case of an emergency, eliminate human errors in manual dispatch, comprehensive security control, surprisingly low-cost |
| 6 | ISO-NE | Anomaly detection (Ref. [56]) | IV.A. IV.B.1).c) IV.C.2) IV.C.4) | Public (AWS) | PaaS | No | Cloud resources usage | Low | Role-based access, MFA, data encryption, HTTPS/TLS | Cluster is highly scalable and cost-effective, cluster configuration is easy, and the workload is low |
| 7 | PGE | Operational efficiency and customer experience improvement (Ref. [57]) | IV.A. IV.B.1).a) IV.B.1).d) IV.C.2) IV.C.3) IV.C.4) | Public (AWS) | PaaS IaaS | Existing software license cost | Cloud resources usage for data lake and analytics | Low | Account or OU-based based access, data classification and cybersecurity scrutiny | Ability to view/analyze both structured and unstructured data, reduced cost, easy and secure data sharing with any 3rd parties, self-service reporting, data traceability, agile development of ML models |
| 8, 9 | Centrica[8], AutoGrid[9] | DER aggregation and management (Ref. [58] [59]) | IV.A. IV.B.1).a) IV.B.1).b) IV.C.3) IV.C.4) | Public (AWS) | IaaS[8] PaaS[8,9] | No | Cloud resources used for data processing, storage, analytics and networking | Much lower than an on-premises solution | MFA[8,9], identity-based access[8,9], data encryption[8,9], activity logging[8,9], 3rd party testing and monitoring of vulnerabilities[8,9], VPNs on IPsec[9], mutual TLS[9], intrusion | Shorter development cycle, better system scalability, simpler DevOps management |

| | | | | | | | | | |
|---|---|---|---|---|---|---|---|---|---|
| | | | | | | | | detection[9], role-based API-level access control[9] | |
| 10 | AEMO | Market settlement (Ref. [60]) | IV.A. IV.B.1).c) IV.C.3) IV.C.4) | Public (Azure) | PaaS CaaS | No | Cloud resources usage | Low, work is automated | Unknown at the time of writing | The market settlement reduced from 30-min to 5-min blocks with massive data management capability |
| 11, 12 | MISO[11], NRECA[12] | Collaborative system modeling and hybrid simulation (Ref. [61]) | IV.A. IV.B.1).a) IV.B.1).b) IV.B.1).d) IV.C.3) IV.C.4) | Public (AWS) | SaaS[11] PaaS[12] IaaS[12] | No | subscription fee[11], Cloud resources usage[12] | Low, fully managed by the vendor | Access control[11,12], incident response[11], System and communication protection[11,12], encryption for data at rest[11,12] | Scalable infrastructure[11,12], modeling consistency[11,12], reduced IT effort[11,12], integrated security, easy to access users' model[12] |
| 13 | IncSys/PowerData | Coordinated system operation drill | IV.A. IV.B.1).c) IV.B.1).d) IV.C.3) IV.C.4) | Public or Private | SaaS | Depends on cloud type | cloud resources usage | Different levels of effort | Firewalls, security groups, access control, intrusion detection, logging and monitoring, data encryption, TLS | Quick availability for customer, scalable and fault-tolerant to support mission-critical drills, Access from anywhere |

*A. Cloud Applications in Grid Planning:*

*1) Planning Studies*

We first consider three use-cases of cloud for planning studies from New York ISO, ISO New-England and Omaha Public Power District.

**New York ISO** (NYISO) –

NYISO started using an on-premises cluster for system planning studies in 2012. As the hardware aged, and workload increased, it became necessary to procure more resources to obtain results in a reasonable time. However, the cost of replacing the entire on-premises system with up-to-date hardware was considered cost prohibitive. Instead, transitioning to the public cloud service was found to be more feasible since the balance of Total Cost of Ownership (TCO) spread out over the effective lifetime of that server until it is decommissioned is comparatively higher. Consequently, NYISO adopted a cloud solution in 2017, which phased into production in 2018.

NYISO chose AWS as the cloud provider and opted for IaaS to build its cloud solution. The solution integrated Microsoft HPC as an instance manager and job dispatcher with AWS EC2 service to scale out the virtual machines via EC2 for HPC jobs. The architecture of this cloud-based HPC platform is illustrated in Fig. 9.

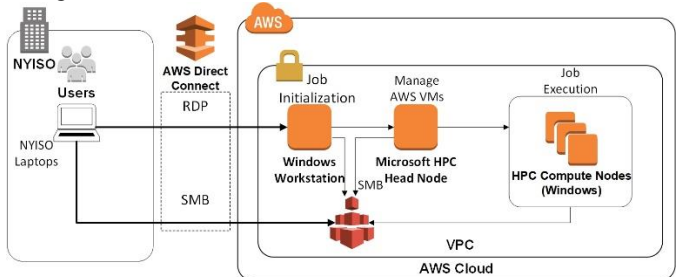

Fig. 9. NYISO HPC Cloud Architecture

As Fig. 9 shows, the simulation jobs are submitted by NYISO engineers to the HPC cluster head node on AWS via a file sharing service which allows on-premises servers to access cloud storage through a dedicated network connection: AWS Direct Connect. These jobs are then dispatched to multiple virtual machines spun up by EC2 service, which runs the job in parallel. The simulation results are sent back to the file system mounted to NYISO's corporate network for users' download. At the time of writing, NYISO's cloud HPC platform can support planning studies in three applications: GE MAPS, GE MARS and PowerGEM TARA.

NYISO uses data encryption and access control to keep its cloud operation secure. All virtual resources acquired from AWS are encrypted by NYISO. End-to-end encryption is used for data transfer between NYISO and AWS data centers. NYISO also enforces strong credentials and high privileges to access the sensitive data. The identities and accesses are controlled by AWS adepts via IAM service.

With this cloud technology, NYISO has achieved significant cost savings. It has also achieved higher scalability to meet various computing profiles and a remarkable reduction in study run-time and job queueing. Over 1200 compute nodes are available to NYISO on cloud when a job is submitted to NYISO's HPC platform, and additional cores can be requested as needed. The job runtime of an exemplary task is 40 minutes on the cloud compared to 12 hours on a local computer without any parallelization.

**ISO New England** (ISO-NE) –

ISO-NE began using a local compute cluster to facilitate engineers with large-scale simulations in 2005. As was the case with NYISO, ISO-NE migrated to cloud for planning studies in 2013 and today is a pioneer of cloud adoption in industry. Like NYISO's solution, ISO-NE's elastic computing platform was also architected on an IaaS model using AWS EC2 service, granting them the maximum flexibility for customization and performance/cost optimization. Fig. 10 shows the architecture overview of the cloud platform.

As shown in Fig. 10, ISO-NE separates the public-facing web server from the back-end servers using both public and private subnets as recommended in Section IV.A.2)a). They adopted a comprehensive security control scheme to protect this platform from being accessed by an unauthorized entity. A multi-layer protection scheme in line with the recommended security control shown in Fig. 4 was applied to the workload. At present, ISO-NE's elastic computing platform hosts three primary power system software applications to support both steady-state and dynamic studies: GE MARS, PowerGEM TARA and Siemens PSS/E. The ISO is also evaluating the possibility of moving large-scale Electro-Magnetic (EMT) simulations to the cloud.

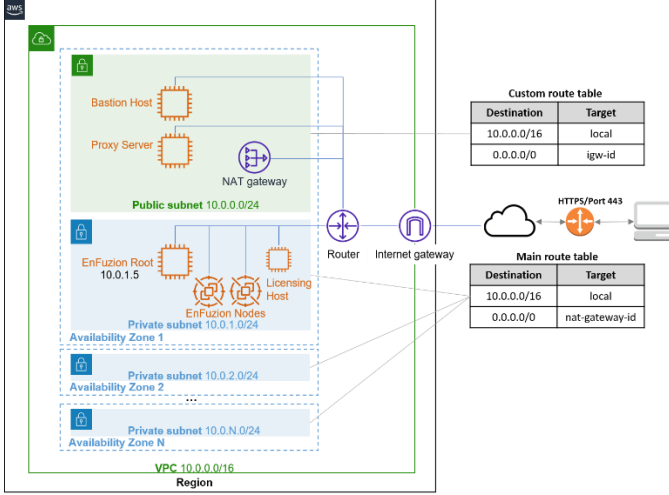

Fig. 10. Architecture overview of ISO-NE's elastic computing platform

ISO-NE's cloud platform shares many similarities with that of NYISO. For example, both platforms are versatile and can accommodate multiple software applications for large-scale power system simulations. The platforms can scale out and down the computing resources based on the demand and utilize unlimited cloud storage. But ISO-NE's cloud platform has some unique features as well, including but not limited to, capability to run instances in multiple AZs to increase fault tolerance, configurable instance type depending on the study's characteristics (if it is compute-intensive or memory-intensive), failed job rescheduling, flexible instance purchasing options (either Spot Instance or On-demand Instance) to strike a balance between cost and performance.

In years of use and enhancement, the platform has demonstrated significant advantages over the traditional on-premises cluster for many planning studies within ISO-NE's workload, including transmission needs assessment and solution study, generation interconnection study, resource adequacy analysis (capacity requirements and tie benefits study), NPCC Bulk Power System (BPS) test, and Forward Capacity Auction (FCA) delist study [18]. A summary of these studies on the platform is given in TABLE VII.

TABLE VII
PLANNING STUDIES THAT RUN ON ISO-NE'S ELASTIC COMPUTING PLATFORM

| Study Type | No. of Jobs | No. of Nodes Used | Nodes Uptime | Cost ($) | Time spent on cloud vs. PC | Time spent on cloud vs. on-prem cluster |
|---|---|---|---|---|---|---|
| N-1-1 contingency analysis | $10^2 \sim 10^4$ | 10 ~ 20 | 1h ~ 6h | $10^1 \sim 10^2$ | ~ 30 times faster | ~ 10 times faster |
| FCA delist study | $10^3 \sim 10^4$ | 100 | < 1h* | 100 ~ 200 | N/A* | N/A* |
| NPCC BPS Test | $10^3 \sim 10^4$ | 10 ~ 20 | 1h ~ 12h | $10^1 \sim 10^2$ | ~ 30 times faster | ~ 10 times faster |
| Demand curve study | $10^2 \sim 10^3$ | 5 ~ 20 | 3h ~ 5h | $10^1 \sim 10^2$ | ~ 40 times faster | ~ 15 times faster |
| Tie benefit Study | 50 ~ $10^3$ | 5 ~ 20 | 1h ~ 5h | 5 ~ 50 | 12 times faster | N/A* |

One example of cloud superiority is visible in FCA delist study. Prior to cloud migration, ISO-NE had never managed to run all N-1-1 scenarios in a FCA delist study as the study is a strictly time-constrained job. The delist study allows ISO-NE to evaluate if there are any reliability issues in case of N-1-1 contingencies with the units' delist bids considered. The ISO must finish the analysis and post the auction results no later than the second day after the auction is closed. With access to unlimited resources on the cloud and the capability to scale out almost instantly, ISO-NE has been able to successfully perform a comprehensive assessment of all possible combinations of contingencies in the study within a tight time window since 2019.

**Omaha Public Power District** (OPPD) -

OPPD began to use HPC with an on-premises compute cluster consisting of 448 usable cores across 14 Windows servers (Intel Xeon 8C and 12C) in total. With this cluster, a typical P6 run [62] with 50 cases, which takes about 3 hours per case, only needed a total of 6 hours to complete (~25x speed improvement) with 16 cores allocated to each case. Due to the success of the local cluster in terms of time reduction, OPPD started trials on cloud computing.

The work was done on a single-tenant virtual machine in Microsoft Azure Government Cloud. The Site-to-Site VPN was established to secure the connection. Azure Active Directory was used to manage identity and access in the cloud. In a series of cloud attempts, OPPD explored different request parameters, including scaling strategies (both vertical and horizontal), number of jobs, and number of cores on a job, to evaluate the speed and cost benefits that can be brought by the cloud. After these cloud trials, they concluded that cloud infrastructure improves overall performance of computing by giving them more scaling flexibility, higher resource availability and requiring less maintenance effort. OPPD aims to move to a hybrid cloud model in the future by shifting the variable portion of demand into the public cloud environment while keeping the fixed demand in house to maximize their benefits through cloud while simultaneously taking full advantage of the existing on-premises infrastructure. Such a hybrid setup, as discussed in section IV.A.1)a), is probably the best way forward for the power system companies that are beginning to migrate to cloud as it optimally balances the portfolio of capital investment and O&M spending.

*B. Cloud Applications in Grid Operation*

Now we consider various use cases of cloud technology for grid operation.

*1) Load Forecasting*

**ISO New England** (ISO-NE) –

Load forecasting is a typical power system engineering problem that heavily uses data-driven approaches for modeling and analysis. The complexity of load forecast and its need to process large historical datasets to build high-fidelity models requires high computational power as well as state-of-the-art data-driven approaches. Cloud computing is a perfect resource for both these needs.

ISO New England is also borrowing power from cloud computing to expedite the development of an enhanced short-term load forecasting (STLF) program based on machine learning (ML) approaches.

ISO-NE's current STLF tool, as a built-in module on EMS, predicts the future 4-hour system demand every 5 minutes to assist operators with the real-time operation. The tool employs a Similar Day method to search the best match of the forecast horizon in the past 7 days considering many attributes such as weather, day of the week, season, actual system load of the historical day, etc. The tool can meet the control room's basic needs for STLF, but it requires operators to report several inputs when none of the past 7 days is a good match. It has been observed that the forecasted load demands derived from the ill-matched "similar day" were inaccurate, making it a "garbage in garbage out" box.

In order to speed up the application of newer data-driven methods to improve the accuracy of STLF, ISO-NE adopted a PaaS solution on AWS to streamline the development and test process. The PaaS solution, named SageMaker, did a lot of heavy lifting to provide infrastructure and allows ISO-NE developers to focus on their own use cases. With this platform, ISO-NE data scientists quickly tried out different ML methodologies to develop new STLF algorithms, based on decision tree, random forest, support vector machine (SVM) and K-means clustering. These new ML-based approaches complement the current STLF tool that is based on a single method. With this approach, the effort needed for server provisioning, dependencies installation, environment configuration was waived and the time spent on data preparation, model building, training and validation were greatly reduced. The well-trained model artifacts, if needed, were stored in a special format and deployed in the on-premises environment to connect with the EMS for security compliance. Fig. 11 shows the workflow of ISO-NE's cloud-based ML model development process.

As a public cloud offering, this ML development environment has seamless integration with security control services. Following the best practices recommended by AWS, ISO-NE activates access control, data encryption, user/API activity logging and MFA to mitigate the potential security risk involved with other ML tasks, even though the data used in the current case, system load and weather data, are not confidential.

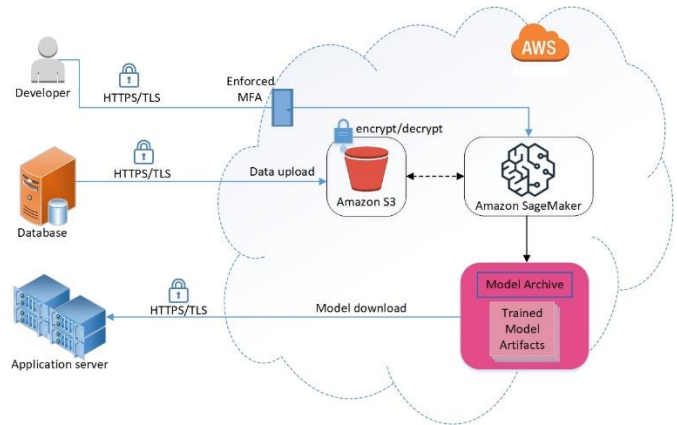

Fig. 11. Schematic diagram of ISO-NE's ML model development flow

*2) Wide-area Monitoring and Data Sharing*

**ISO New England** (ISO-NE)/**New York Power Authority** (NYPA) –

A collaborative, easy-to-share, wide-area situational awareness system was implemented in the cloud through a joint effort by ISO-NE, NYPA, Cornell University and Washington State University [63].

In this system, the synchrophasor data are streamed from Phasor Data Concentrators (PDCs) at each system operator to the cloud-based data relay and then rerouted to various applications that are hosted in the cloud, such as Regional Data Repository and Real-time Phasor State Estimator (SE). The state estimator performs an interconnection-wide, time-synchronized assessment of the voltage magnitudes, phase angles, and transmission line flows. It also provides common visualization displays for real-time collaboration among the participating grid operators. The grid operators who subscribe to this platform can access each other's PMU data as well as retrieve the historical data from the central repository.

The cloud deployment of the wide-area monitoring and data sharing platform is called "GridCloud". Its implementation architecture can be found in [63]. The GridCloud system was built on AWS, particularly VPC and EC2 service, with the duplicated workloads in different Regions to achieve high fault tolerance and optimal performance (e.g., reduce the network latency). In this implementation, PMU data were streamed from the distribution points at ISO-NE and Cornell via encrypted secure shell (SSH) tunnels to two redundant Amazon Virtual Private Clouds (VPCs) in Virginia and Oregon. Each data center had 13 cloud instances, with a total average cost of $2.47 per hour. The cloud instances were managed by Cornell's open-source software CloudMake and VSync. Besides the use of encrypted SSH to protect the data in transit, the data at rest was also encrypted using AES-256 algorithm. The encryption did add latency, but it remained within satisfactory range. The data consistency between the PMU data producer and the final application (SE in this case) was verified against each other in both data centers, catering to the strong need of data consistency by grid operators. Data consistency ensures that different viewers see the same data, and any updates to grid state are promptly evident; otherwise, grid operators might take actions that are detrimental to the grid operation.

GridCloud demonstrated that the complex smart grid applications with strong requirements on security, consistency, and timeliness could be supported on the public cloud. The system used redundancy and replication to mask the various sources of delay and jitter and helped to manage tamper-proof real-time archival storage of the collected data. Moreover, the overall cost of the entire system setup, which includes the cryptographic security, was surprisingly low.

*3) Backup Control and Emergency Generation Dispatch*

**ISO New England** (ISO-NE) –

One of the important functions that a system operator performs is balancing load and generation in real time. An operator which assumes such a role is also called Balancing Authority (BA). Under normal situations, the area balancing functionality is mainly achieved through Automatic Generation Control (AGC) and Unit Dispatch System (UDS). AGC is a critical component of the EMS. It calculates Area Control Error (ACE) based on SCADA measurements of tie-line flows and frequencies, computes and sends AGC set points to the participating units every 4 seconds. UDS runs periodically at an interval, usually every 10 minutes, with a look-ahead time window to develop a security-constrained economic solution and the Desired Dispatch Points (DDPs) for all dispatchable resources. When an abnormal circumstance occurs, e.g., AGC fails to work due to EMS shutdown, UDS out of service due to communication network failure, operators have to manually dispatch generators over secure phone calls for area balancing. Unfortunately, such a dispatch procedure involves significant overhead and manual input, making the system operation performance vulnerable to human errors. Besides, it is also progressively difficult to dispatch multiple generators verbally in a short interval to balance the system. Additionally, operators may also have to deal with unusual times when the facilities are inaccessible, like COVID-19. Since not all essential applications for generation dispatch allow remote access, how to continue operating the grid during such scenarios is of particular interest to grid operators. To provide a backup control for area balancing in the event of such emergencies, ISO-NE developed a cloud-hosted emergency generation dispatch solution, which is based on serverless computing. ISO-NE evaluated the method amid COVID-19 pandemic and found it to work as expected.

The solution's serverless architecture has been illustrated in [27]. The entire platform is built upon AWS with its serverless service – Amazon Lambda as the cornerstone. Building upon Amazon Lambda, the platform also integrates other Amazon services for data ingestion, storage and visualization. Whenever an abnormal event occurs that leads to area balancing dysfunction, the operators are required to raise an emergency alarm as required by the current operating procedure. This alarm is used to trigger a periodic retrieval of PMU measurements including tie-line flows, frequencies on key buses, and active power outputs of the PMU-monitored generators, from the PMU database. In the meanwhile, other data such as unit ramp rates, incremental energy offers, operating limits and regulation limits will be pulled from the market database (if that database also fails, the last successfully retrieved dataset is used). The data are then securely sent to an S3 bucket (an Amazon cloud storage service) and encrypted. The data upload operation then triggers a Lambda function to start calculating ACE and DDPs for the dispatchable generators. The calculation results are sent to another S3 bucket where a static website is hosted with the data visualized through Data-Driven Document display technology (d3.js) [64]. The results are also written to DynamoDB, a NoSQL database service on AWS, for archiving purposes.

The BA and the generator operators can visit their respective webpages to visualize the dispatch information through HTTPS links in terms of signed Uniform Resource Locators (URLs) after they authenticate themselves with a pre-defined user pool in Amazon Cognito. The operators at the BA side are allowed to manually override the advised DDPs calculated by the Lambda function if they think these values are unreasonable, whereas the generator operators can acknowledge the advised DDPs or give a reason to decline it. Similarly, generator operators can initiate a request to change the Unit Control Mode (UCM) of a unit and BA can confirm this change. These user requests are completed by API calls through Amazon API Gateway. All the operation activities are logged in the DynamoDB database for auditing and responsibility-tracking purposes as a replacement for phone conversation recording.

To protect data in transit, HTTPS with Transport Layer Security protocol is enforced. As for protection of data at rest, the server-side encryption with AWS KMS managed keys is used. The data keys rotation is activated through KMS, and the master key is also manually rotated periodically for enhanced security.

With the aid of this cloud-based solution, the potential human errors due to the use of phone conversations for manual verbal dispatch can be avoided. Each operation activity done through the platform is written to a database in a clearly defined data format, which makes auditing and responsibility tracking much easier. More importantly, the solution also enables the operators to monitor and control the grid even when they need to work remotely away from the control room as long as they have access to the Internet connection through an encrypted SSH tunnel. Thanks to adoption of the event-driven Function-as-a-Service (FaaS) model, there is no charge at the ISO under normal operation conditions. The overall cost of this solution is extremely low considering the infrequent occurrence of an emergency.

*4) Anomaly Detection*

**ISO New England** (ISO-NE) –

Power utilities gather various types of data from day-to-day operations. As power systems modernize, many utilities/ISOs are unable to analyze the sheer amount of new information the grid sensor collects. This inability to analyze the massive quantity of data has been further aggravated as the number of PMU devices grow [56]. The system measurements, especially the PMU data, contain a substantial amount of valuable information that reflects the system conditions. Mining of these data to identify those observations that deviate from the system's normal behavior could help power utilities uncover the hidden patterns or issues in the system. For that goal, there is a need to perform analysis on massive datasets with big data analytics methods to obtain useful information about the system. To be of practical value, however; these methods have to be much faster than the traditional data processing tools.

One use case which has become ubiquitous in ISOs is the efficient identification of frequency excursions. Frequency excursion is a direct result of an imbalance between the electrical load and generation. To maintain frequency stability, it is necessary to effectively control the frequency within a defined range. The *Frequency Response Initiative Report* by NERC recommends all turbine-generators to be equipped with governors so as to provide immediate and sustained response to abnormal frequency excursions. A specific frequency response performance requirement is also given in this report for power plants to comply with: "Governors should, at a minimum, be fully responsive to frequency deviations exceeding ±0.036 Hz (±36 mHz)" [65]. Balancing Authorities (BAs) or their equivalent entities' responsibility is to monitor, measure, and improve system total frequency response as well as audit governor response of larger power plants. Currently many BAs rely on FNET/GridEye [66] to keep informed of system disturbance events. FNET collects data from over 300 Frequency Disturbance Recorders (FDRs) deployed in the North America power grid. The accuracy of the frequency measurement obtained from the FDR can reach ±0.0005 Hz. When a frequency measurement deviates from the system nominal frequency (60 Hz) significantly in a short period, FNET is able to provide real-time alerts via emails. However, FNET is insensitive to slow frequency changes (i.e., small df/dt) and might miss the events when the frequency deviation is significant, yet the rate of frequency change is small. Another concern is utility engineers may not always pay attention to the alert notifications in a timely manner. When they start looking back on the frequency excursion events to analyze governor response, it is not easy to categorize these alerts by their criticality and prioritize them for study. In order to meet NERC's requirement for BAs to submit their annual report of compliance with frequency response performance specified in [65], ISO-NE built a scalable big data analytics platform on AWS to enable fast and accurate identification of large frequency deviation events from historical dataset.

As shown in Fig. 12, the core service of the big data platform is Amazon EMR (Elastic MapReduce), which is used to create an elastic cluster with as many nodes (EC2 instances) as needed based on the projection of data growth. EMR is seamlessly integrated with Amazon S3 service to directly access data stored in permitted S3 bucket as if it were a file system like Hadoop Distributed File System (HDFS) [67]. The EMR cluster natively supports most up-to-date data analytics applications, including big data analytics engines such as Apache Spark and Hadoop, machine learning platforms like Tensorflow, data warehouse and interfaces like Apache Hive, and interactive notebook tools like Jupyter and Zeppelin.

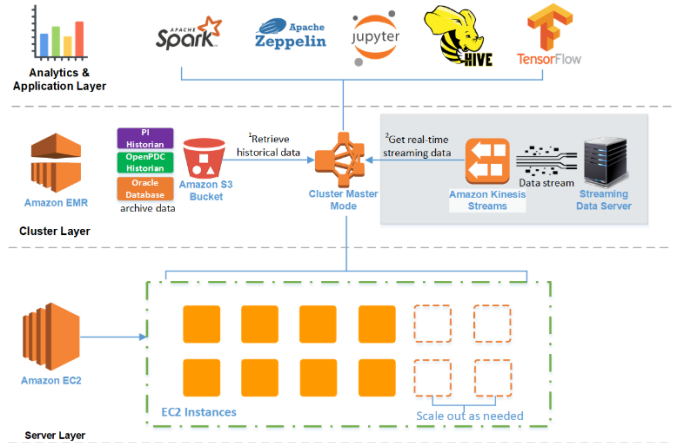

Fig. 12. ISO-NE's scalable big data analytics platform

From the security control perspective, the entire platform adopts a Role-based Access Control (RBAC, refers to section IV.A.2)b) for details). The users/data analysts whoever intend to access the notebook environment are required to login using MFA. The platform is highly cost-effective because the cluster is scalable and compute nodes are launched as Spot Instances. The cluster configuration is a no-brainer. The developer can easily modify the parameters to size the cluster and work with a massive volume of data. There are two ways to acquire the PMU data from ISO-NE, one is through batch upload and other through real-time streaming. For the former approach, the historical synchrophasor data in archives are uploaded to an S3 bucket periodically through either AWS Command Line Interface or AWS API. For the streaming method, Amazon Kinesis is used to ingest the real-time PMU data streams (this part is still under development). Tests show that the platform could help the engineers quickly identify the time of large frequency excursions occurrence when they need to evaluate if the governor response satisfies NERC's requirement [56].

*5) Operational Efficiency and Customer Experience Improvement*

**Portland General Electric** (PGE) –

The utilities employ the smart meter data as well as the customers' input to identify the problems quickly and accurately in their service territory. The comprehensive analytics helps them make decisions about when to take preventive measures after a power pole health inspection or estimate the service restoration time based on the statistics of historical records. These contribute to the improvement of the utilities' operational efficiency and customer experience. PGE is one of the utilities which pioneered using cloud to accomplish these goals.

PGE moved most on-premises databases that spread across the company to a centralized data lake in the cloud, which enabled PGE to perform advanced analytics on massive amount of heterogeneous data and roll out multiple new digital channels such as Intelligent Voice Assistant (IVA), website improvement, mobile app, and proactive outage notification. The cloud adoption at PGE is strongly motivated by the goal to provide data driven insights for strategic planning and to avoid monolithic architecture from the solution design perspective. The direct outcome of this change is better customer engagement and higher operational efficiency.

PGE developed two platforms for cloud-based analytics: a fine-grained API microservices platform for database migration of the on-premises customer outage management system to reduce customer friction and a data lake for performing advanced analytics and machine learning tasks. The overview of the database migration solution and the architecture of the data lake can be found in [57]. Through the cloud-hosted customer outage management solution, PGE customers are provided an omni-channel experience to log in from any of the channels, the mobile or web, to perform a function like making a payment, viewing or report an outage, or to manage their products and programs. It is said that the platform helps PGE reduce the planned outage time of the customer support service by 1/10 and hence increase the customer satisfaction. On the other hand, the data lake helps PGE sunset some of their on-premises data marts/warehouses and is expected to save more than 3 million dollars in the next couple of years. It gives PGE the ability to isolate and increase compute power independent of storage. One highlighted example is sharing of AMI billing data with the distributed energy resources partners, which can be done in a matter of minutes in comparison to a few days when done on-premises systems. Another example is the reduction in time (reduced by 168 hours/year) for customers to report their energy usage with The Environmental Protection Agency (EPA) to track their greenhouse gas emissions.

Furthermore, with access to ML models that can be quickly prototyped and deployed to seamlessly integrate with the data lake, the data scientists at PGE are able to improve the accuracy of Estimated Outage Restoration Time (EORT) from 24% to 59%, thus reducing about 700,000 unnecessary messages sent to the customers for EORT updates at the onset of outages. The accuracy improvement has also lowered the effort of field crews on updating the outage information while they are working hard on restoring the customers' power supply.

Moreover, PGE also harnesses the heterogeneous data that are collected within the data lake, such as AMI meter measurements, photovoltaic (PV) panel capacity and weather data to predict the BTM solar generation for better distributed resource planning (DRP) [68]. This helps properly plan out the grid assets to meet the customer's demand when using cheaper clean energy.

PGE applies a comprehensive cybersecurity audit on the cloud platform and any data used on it. They discriminate data based on their confidentiality, classify their accounts (Organization Units, or OU) in terms of purpose of usage, e.g., DEV, TEST, and PROD, and implement identity-based access to these accounts. The maintenance effort of this solution can be fully automated through the cloud and is claimed to be less time intensive. The data management team, which consists of 7 employees, was overwhelmed by the on-premises data warehouse maintenance work. Now they can maintain both the on-premises and the cloud workload effortlessly.

### C. Cloud Applications for Other Business Needs

#### 1) DER Aggregation and Management

As a prevailing solution for DER aggregation and unified management, Virtual Power Plant (VPP) has drawn a lot of attention. A VPP is a centralized platform that takes advantage of information and communication technologies (ICTs) and Internet of things (IoT) devices to aggregate the capacities of heterogeneous DERs to form "a coalition of heterogeneous DERs". It acts as an intermediary between DERs and the wholesale electricity market and trades energy on behalf of DER owners who by themselves are unable to participate in that market. VPP is a cloud-native platform, which means it utilizes cloud computing to "build and run scalable applications in a cloud environment". Centrica and AutoGrid are two examples whose VPP solutions are built on public cloud services.

**Centrica –**

Centrica has been a solution provider for distributed assets monitoring and management for 10 years. Centric's VPP platform has 1.7GW of DERs under management. The product, named FlexPond, delivers an industry-leading reliability to ancillary service markets across Germany, UK, Belgium and France. The solution architecture of FlexPond can be found in [58].

Centrica adopted AWS as the cloud provider and built the solution using services such as AWS IoT Core and AWS IoT Greengrass. IoT Core lets connected devices interact with cloud applications and other devices securely and effortlessly. IoT Core can support billions of devices and trillions of messages and can process and route those messages to AWS endpoints and to other devices reliably and securely. IoT Greengrass seamlessly extends AWS to devices so they can act locally on the data they generate, while still using the cloud for management, analytics, and durable storage. Through the public cloud offerings, Centrica was able to shorten the product development time, increase the system scalability and simplify the DevOps flow. According to Centrica, much lower maintenance effort is needed in this cloud solution compared to the on-premises solution.

Centrica secures all communications, stored data and continuously monitors the system integrity. The VPP platform is not accessible from the public Internet. Any authorized access requires two steps: login to a private network and two-factor authentication. All communication between the users and the platform is encrypted with strong cryptography. The data associated with the DERs management are stored in secure, ISO 27001-certified data centers. The access to the data is strictly limited based on the user identity and corresponding access rights. The platform uses cloud services to log, monitor and retain account activity related to actions on data. Besides, it also employs third party expertise for testing and monitoring of vulnerabilities in the system.

A case study at Terhills in Belgium has shown that Centrica's VPP is able to aggregate a diverse range of heterogeneous DERs from generation (combined heat and power), storage (batteries), micro-production (wind and solar), and flexible loads (residential, industrial and commercial).

IEEE TRANSACTIONS ON SMART GRID 27Centrica launched a 32-megawatt virtual power plant, with a distribution grid connected to the 18.2-megawatt Tesla Powerpack storage system in the Terhills project. The installed VPP has been providing primary frequency response to the Belgian TSO since April 2018 and stacking value with additional participation in the real time balancing market. This battery project is unique because it is included in a larger flexibility portfolio, which results in a 1.5 times higher revenue stream for the battery, compared to the base case where the battery is monetized standalone.

**AutoGrid** –

AutoGrid VPP aggregates customer-owned flexible storage, distributed generation and demand-side resources to monetize resources in multiple energy markets and turn them into cash generators. With state-of-the-art forecasting and asset-optimization capabilities, AutoGrid VPP allows utilities and aggregators to create additional flexible capacity and extract maximum value from flexible resources in lucrative markets around the globe.

AutoGrid also chose AWS as the CSP to build its VPP solution upon a vast range of services and features to secure the workload and demonstrate the compliance with security standards. The VPP platform [69] is hosted on a secure logical network and access to it is limited to ports from known IPs using security groups and network access control lists. AWS IAM is used for fine-grained, API-level control of remote access for administration of cloud assets. The software itself runs on a highly scalable and resilient Kubernetes architecture on Amazon Elastic Kubernetes Service (Amazon EKS) within the protection of an Amazon VPC. All data in Amazon S3 is encrypted with keys that are managed by the AWS KMS, which implements FIPS [70] compliant modules and uses AWS IAM to control access to keys. For logging and monitoring, the solution uses AWS CloudTrail and Amazon CloudWatch for alarms and notifications that help with incident response. For network intrusion detection, AutoGrid uses Amazon GuardDuty.

AutoGrid practices cybersecurity from the ground up and builds their grid products baked in with cybersecurity capabilities and controls. The security challenges mentioned above as well as demonstrating compliance requirements are addressed by appropriate security controls implemented by AutoGrid in collaboration with AWS cloud infrastructure. Like Centrica, these security controls are also verified and attested by an independent third-party auditor.

*2) Market Settlement*

**Australia Energy Market Operator** (AEMO) –

AEMO is Australia's independent energy markets and power systems operator, and system planner. One of its principal responsibilities is settlement of the $16 billion-plus National Energy Market (NEM) which connects the grids of eastern and southern Australia states and territories to create a wholesale energy market. Retailers and wholesale consumers pay AEMO for the electricity they use, and AEMO then pays the generators.

Currently, the National Energy Market is dispatched in five-minute slots – but only settled every 30 minutes due to limitations in its ability to segment the data. AEMO has been working with Microsoft and partner Tata Consultancy Services to transition from a legacy settlement system to its big data Metering Data Management solution built on Microsoft Azure cloud. Cosmos DB is the main data store, leveraging Azure Kubernetes Services for the application and runtime layers. The ability to easily manage huge amounts of data allows settlement in five-minute blocks for AEMO. The switch to five-minute settlement (5MS), scheduled for October 2021, aligns the dispatch and settlement times and is expected to remove a number of potential market barriers for renewable energy providers for bidding and dispatching as well as encourage additional innovation. The solution should provide a stronger incentive for market participants to respond to the rapidly changing dynamics of the electricity market [71].

*3) Collaborative System Modeling and Hybrid Simulation*

**Midcontinent ISO** (MISO) –

The trifold role of MISO requires accurate power system models for operations, market, and planning activities, which becomes increasingly challenging as volume of model data increases and frequency of change accelerates due to integration of distributed renewable resources and dynamic model refinement.

Currently, MISO updates the operation models on a quarterly cycle. However, the members of MISO are expecting more frequent model updates. MISO has multiple disjointed customer requests for the same model data in operation and planning horizons. The ISO gets model data multiple times in different formats. As a result of which, there is a fair chance that discrepancies exist between the online and offline model when processes are out of synchronism, creating a barrier for model management with continuous increases in volume and frequency of change of modeling data. Power system model errors inevitably occur due to manually intensive and duplicative processes, resulting in inefficient use of human resources and underutilization of energy resources in the generation fleet. By means of investing in transformative modeling processes that leverage the state-of-the-art cloud technology, MISO can ensure the power system models are accurate, synchronized and consistently updated across ISO activities, and transparently provided to its members in shorter update cycles and with reduced errors.

The new solution that MISO is going to adopt for network model management (NMM), is built on top of AWS public cloud services. The solution is delivered to MISO from its vendor Siemens via a SaaS subscription, bringing the ISO a scalable infrastructure, consistent application performance, reduced need for IT support, as well as security assurance from both the cloud provider and the vendor. The SaaS model shifts all IT-related maintenance effort to the vendor, while MISO only needs to grant the use of this system to its members and provide necessary training to them. The system has been scheduled to go into production in late 2021.

Although users have concern over data security, the vendor has a vested interest in providing the highest level of data security. The NMM system has changed logs to support audits of access to model data whenever necessary. On the other hand, MISO takes security of systems in the cloud seriously. The security requirements that have to be met cover the following areas: Access Control, Audit and Accountability, Awareness and Training, Configuration Management, Contingency Planning, Identification and Authentication, Incident Response, Maintenance, Media Protection, Personnel Security, Physical and Environmental Protection, Planning, Risk Assessment, Security Assessment and Authorization, System and Communication Protection, System and Information Integrity, System and Services Acquisition.

**National Electric Cooperative Association** (NRECA) –

NRECA is a national non-profit service organization that serves over 890 rural electric utilities. NRECA offers Open Modeling Framework (OMF.coop) at no cost to its member utilities via a combination of internal and federal funding. The OMF.coop is a software development effort led by NRECA with a goal of making advanced power systems models usable in the electric cooperative community. OMF.coop addresses the lack of versatile modeling tools that would enable utilities to evaluate smart grid components using real-world data prior to purchase. NRECA has adapted the OMF to meet utilities' need for a tool that can combine and analyze data resulting from the integration of new renewable resources such as wind and solar, as well as other distributed energy resources. This enhanced modeling tool can support co-op investment decision-making by modeling the cost and benefits and incorporating engineering, weather, financial and other data specific to the co-op.

OMF.coop solution is hosted in the AWS public cloud and uses a combination of IaaS (Amazon EC2) and PaaS (Amazon SES, Amazon S3) services. The on-demand features of the cloud allow OMF.coop to easily scale the solution based on the user growth. Exposing the software via a web interface backed by cloud infrastructure provides a better outreach and support to the user community without having to access or modify their systems.

OMF.coop software has been in production for a couple of years, most of the maintenance has been automated. While some man-hours are spent per month to verify new code deployments, the overall cost of the cloud-hosted solution is insignificant.

*4) Coordinated System Operation Drill*

**IncSys/PowerData** –

IncSys and PowerData are strategic partners that lead a cooperative effort in developing a software solution to help power system organizations of all sizes train and prepare system operators to ensure the reliability of the bulk power system. Their training software, PowerSimulator, allows NERC entities to test their restoration plans with voltage and frequency response. It allows simulation of the most devastating system events, such as wildfires, hurricanes, tornadoes, ice storms, earthquakes, and gas curtailment due to cold snap. PowerSimulator also helps these organizations to test and develop the proficiency of their system operators on the tasks they perform under normal and emergency operations. The representative users of PowerSimulator include ISO New England, FRCC, PJM, WECC, Central Maine Power, Vermont Electric, NSTAR, National Grid, etc.

PowerSimulator allows users to simulate and train on their own system with a high level of realism. For example, they can start and redispatch generation and see the effect on equipment/path MW, MVAr and MVA flows, as well as bus voltages and angles. They can also control voltage with transformer taps, generator kV set points, shunt capacitors, shunt reactors and SVCs. Operators can develop, train on and test switching orders before they are executed. Every type of bus configuration including main and transfer, breaker and half, double breaker double bus, double bus single breaker, ring bus and single bus single breaker can be modelled in the simulator.

Being a native web application, PowerSimulator is fast and runs in the most modern browsers such as Edge, Chrome and FireFox. The cloud allows for simulation participation from any site with Internet access and a modern browser. Bringing realism to drills and training as participants access the software from their normal place of work using the same communication tools they would in a real contingency. Three-way communication is heavily exercised by multiple NERC entities. All participants work on a real-time interactive session playing their actual roles. The actions of each operator and its impact on the system are seen in real-time.

PowerSimulator is offered to its clients as a SaaS model. It is normally provided on a public cloud where IaaS services such as network, compute, storage and security are utilized to build this cloud-based solution. It can be also hosted in a private cloud or in a virtualization environment through VMWare and Hyper-V. However, moving to the public cloud in this case does bring several benefits over other platforms according to IncSys and PowerData. First, the public cloud allows for quick availability. Single-tenant virtual machines can be easily and quickly set up and brought online to meet growing customer demands, allowing end users to test new add-ons, custom features and other necessary upgrades in an independent test environment without the need to revert these changes later. Secondly, Public cloud is a scalable, reliable and secure infrastructure providing peace of mind over hardware failure during mission critical drills. It allows utility IT departments to focus on maintaining their own EMS and SCADA without being loaded up with training simulator issues. The scalability brought by public cloud allows for more than one hundred users to participate in one drill, creating a sense of realism. Multiple users can participate in the roles they would perform in real life. Additionally, going to the public cloud eliminates the traveling costs of operators who participate in the coordinated drill as training can be done online.

The simulation training workload hosted in the cloud is secured by a multifaceted protection scheme via following the best practices for cloud-level application security. The protection includes: a) minimizing public exposure and attack surface: servers respond to Internet requests only for the authorized service, cloud-vendor firewalls and security groups are used as an additional layer of protection, and access can also be restricted from specific organizations; b) Penetration testing is performed at regular intervals; c) Agents run on each server to

IEEE TRANSACTIONS ON SMART GRID

continually monitor the system and detect possible intrusions; d) System-level logs are transferred to a service that allows for managed retention and review; e) All CEII data is encrypted while at rest, and while in transit between persistent stores, server, and end users; f) Requires modern transport layer security (TLS) for encrypted communication, server identification and client-side authentication; g) Auditing of all automated and manual activities affecting the configuration of cloud resources.

## VI. Conclusion

Cloud computing is a mature technology, which many industries are aggressively adopting to meet their business needs. In the power industry, cloud technology is appealing to an increasing number of practitioners as well. Through cloud-based technologies, power system organizations can realize their digital transformation goals efficiently that cannot be met with on-premises resources alone. They can also adapt to the grid modernization process. However, several bottlenecks to using the cloud in power systems are due to common misconceptions about the technology and concerns about service quality, cost, security, and compliance. This paper aims to quell these misconceptions and provides guidance to overcome these bottlenecks. It also puts forth a body of knowledge to assist regulatory bodies in developing standards and guidelines related to cloud adoption and compliance attestation. With the abovementioned challenges addressed appropriately, the cloud will become a widely accepted technology and play a significant role in digitalizing the power industry.

## VII. Discussion

This paper provides guidance for cloud use and portability specific to the needs of the power industry. However, it should be noted that other generic guidance for cloud use and portability from other critical industries such as government [51], [72], health [73], and finance [74] also can apply to the power industry. Nonetheless, there is a notable difference between the power industry and these other critical industries. While there are specific compliance requirements and programs for these critical industries, no such standardized compliance requirement exists for the power industry. For instance, the Health Insurance Portability and Accountability Act (HIPAA) [73] for healthcare, FedRAMP [72] for government, and the Center for Financial Industry Information Systems (FISC) [74] for financial services serve as compliance programs for these critical industries. They have helped the cloud become a regular solution to address many of their needs and challenges. However, to date power industry has no such standardized compliance program.

## VIII. Acknowledgment

The authors would like to thank a group of Task Force members for providing the facts about how they use cloud-based solutions to address their specific business needs. These contributors are Sung Jun Yoon, Uma Venkatachalam, and Aravind Murugesh from PGE, Xing Wang from Centrica, Scott Harden from Microsoft, Michael Welch from New York ISO, Michael Swan from OPPD, David Pinney from NRECA, David Duebner from MISO, and Chris Mosier from IncSys.